\documentclass[conference]{IEEEtran}

\usepackage{amsmath,amssymb,amsfonts}
\usepackage[ruled,vlined]{algorithm2e}
\usepackage{float}
\usepackage{graphicx}
\usepackage{textcomp}
\usepackage{xcolor}
\usepackage[numbers]{natbib}
\usepackage{hyperref}
\usepackage{placeins}
\usepackage{pdflscape}
\usepackage{qcircuit}
\usepackage{booktabs}
\usepackage{mathtools}
\usepackage{pgfplots}        
\pgfplotsset{compat=newest}  
\usepackage{caption}         
\usepackage{graphicx}        
\usepackage{epstopdf}
\def\ket#1{|{#1}\rangle}

\DeclarePairedDelimiter{\floor}{\lfloor}{\rfloor}

\allowdisplaybreaks 

\def\BibTeX{{\rm B\kern-.05em{\sc i\kern-.025em b}\kern-.08em
    T\kern-.1667em\lower.7ex\hbox{E}\kern-.125emX}}

\begin{document}

\title{Adaptive Genetic Algorithms for Pulse-Level Quantum Error Mitigation}

\author{
    \IEEEauthorblockN{William Aguilar-Calvo}
    \IEEEauthorblockA{
        \textit{School of Software Engineering} \\
        \textit{CENFOTEC University} \\
        San Jos\'e, Costa Rica \\
        waguilarc@ucenfotec.ac.cr}
    \and
    \IEEEauthorblockN{Santiago Núñez-Corrales}
    \IEEEauthorblockA{
        \textit{NCSA, IQUIST} \\
        \textit{University of Illinois Urbana-Champaign} \\
        Urbana IL, USA \\
        nunezco2@illinois.edu}

}

\maketitle
\begin{abstract}
Noise remains a fundamental challenge in quantum computing, significantly affecting pulse fidelity and overall circuit performance. This paper introduces an adaptive algorithm for pulse-level quantum error mitigation, designed to enhance fidelity by dynamically responding to noise conditions without modifying circuit gates. By targeting pulse parameters directly, this method reduces the impact of various noise sources, improving algorithm resilience in quantum circuits. We show the latter by applying our protocol to Grover's and Deutsch-Jozsa algorithms. Experimental results show that this pulse-level strategy provides a flexible and efficient solution for increasing  fidelity during the noisy execution of quantum circuits. Our work contributes to advancements in error mitigation techniques, essential for robust quantum computing.
\end{abstract}


\section{Introduction} \label{sec:introduction}

As quantum computing advances toward practical realization, the pervasiveness of hardware sensitivity to noise sources limits its transformative potential. Quantum computers promise extraordinary capabilities for solving problems otherwise intractable for classical systems, such as factoring large numbers, simulating complex quantum systems, and optimizing combinatorial problems \cite{b1}. Addressing this challenge is imperative for the next leap in computational power, and achieving scales of practical utility enabled by fault-tolerant quantum computing systems.

Contemporary quantum hardware operates within the constraints of Noisy Intermediate-Scale Quantum (NISQ) technology \cite{b2}, where noise introduces significant barriers to the fidelity and scalability of quantum operations. As highlighted in Preskill's foundational work on NISQ \cite{b2}, noise limits the reliability of quantum computations by introducing errors. In turn, reducing error rates prompts innovative methods for error mitigation. Quantum errors arise in various forms including \textit{bit-flip} and \textit{phase-flip} errors which respectively alter the logical state and phase of individual qubits \cite{b3}. Decoherence processes, characterized by amplitude relaxation (\(T_1\)) and phase decoherence (\(T_2\)), degrade quantum information over time \cite{b24}. Multi-qubit systems face \textit{crosstalk}, where unwanted interactions between qubits introduce additional errors during the execution of two- and multi-qubit gates \cite{b25}. These error mechanisms underscore the urgent need for scalable and robust strategies to mitigate noise in quantum systems. Since physical mechanisms responsible for the susceptibility of a quantum system to intrinsic and external noise sources tend to be harder to address, software-based strategies constitute the center of attention in this article.

Traditional Quantum Error Correction (QEC) methods address these challenges by encoding logical qubits into entangled states distributed across multiple physical qubits \cite{b4}. For instance, a single logical qubit state \(|\psi\rangle = \alpha|0\rangle + \beta|1\rangle\) can be transformed into an encoded state:

\[
|\psi\rangle \rightarrow |\psi_L\rangle = \alpha|0_L\rangle + \beta|1_L\rangle,
\]

where \(|0_L\rangle\) and \(|1_L\rangle\) represent logical states encoded across multiple physical qubits. Syndrome measurements identify errors and enable corrective operations to restore \(|\psi_L\rangle\) to its intended state \cite{b4}. While effective, QEC requires significant overhead in qubits and operations, posing challenges for current NISQ hardware. Given these limitations, alternative methods that can effectively operate within NISQ constraints are essential.

Pulse-level control offers a direct means to mitigate noise by manipulating the Hamiltonian of the quantum system, bypassing the abstractions of gate-level programming \cite{b7}. The time-dependent Hamiltonian \(H(t)\) governs the dynamics of the quantum system and is expressed as:

\[
H(t) = H_0 + \sum_k u_k(t) H_k,
\]

where \(H_0\) represents the intrinsic Hamiltonian of the system, \(H_k\) are control Hamiltonians corresponding to specific operations, and \(u_k(t)\) are time-dependent control parameters applied via external pulses \cite{b21}. By precisely adjusting \(u_k(t)\), pulse-level control allows the system to steer towards desired quantum states with high fidelity, addressing noise at its source.

Advances in pulse optimization techniques such as Gradient Ascent Pulse Engineering (GRAPE) and Chopped Random Basis (CRAB) algorithms \cite{b11,b12, Shammah_2024} have improved the ability to design high-fidelity control pulses \cite{b9}. These methods optimize control functions \(u_k(t)\) to achieve desired quantum operations while adjusting for system dynamics \cite{b9,b12}. However, these are typically applied offline and may not easily adapt to real-time noise variations in quantum systems \cite{b9,b11}. Despite these advances, a lack of adaptive methods that can dynamically mitigate noise without increasing circuit complexity remains. This gap hinders the practical implementation of quantum algorithms on NISQ devices.

Adaptive optimization techniques, particularly Adaptive Genetic Algorithms (AGAs) \cite{b8}, offer promising avenues for addressing dynamic noise in quantum systems by adjusting parameters in real time. By leveraging feedback from the quantum system, AGAs can dynamically adjust pulse parameters to mitigate noise effects, increasing gate fidelities without modifying the original structure or composition of the entire circuit. Moreover, the application of AGAs to quantum error mitigation exemplifies a case in which the principled application of automated learning methods contributes to a better separation of concerns between levels of abstraction across the quantum stack \cite{di2024abstraction}: the task of achieving greater fidelities differs conceptually from that of executing a quantum algorithm, and (ideally) should not be a visible concern to quantum programmers.

\subsection{Contributions} \label{sec:contributions}

This paper introduces a novel adaptive pulse-level quantum error mitigation algorithm comprising three main advances:

\begin{enumerate}
    \item \textbf{Adaptive Pulse Parameter Optimization:} A new algorithm dynamically adjusts pulse parameters in response to real-time noise measurements, thereby enhancing its responsiveness to fluctuating noise conditions within quantum systems.
    
    \item \textbf{Enhanced Fidelity Without Circuit Alteration:} The algorithm achieves substantial fidelity improvements in quantum algorithms without requiring modifications to the original circuits, thereby preserving their design and intent.
    
    \item \textbf{Empirical Validation on Benchmark Algorithms:} The efficacy of the proposed method is empirically demonstrated through enhanced performance metrics in two benchmark quantum algorithms, specifically Grover's and Deutsch-Jozsa algorithms.
\end{enumerate}

Collectively, our contributions underscore the potential of evolutionary strategies to refine pulse-level control and achieve error mitigation. These also suggest an effective methodology to enhance the reliability of quantum computations on Noisy Intermediate-Scale Quantum (NISQ) devices. By leveraging adaptive optimization techniques, our methodology effectively navigates the noise of current quantum systems, thereby enabling more robust and accurate computational results.

\subsection{Organization of the Paper} \label{sec:organization}

The remainder of this paper is structured as follows: \textbf{Section~\ref{sec:background}} provides the necessary theoretical and technical background encompassing noise modeling in quantum computing, state-of-the-art quantum error modeling, quantum error mitigation and correction techniques, and a brief introduction to QuTiP. \textbf{Section~\ref{sec:methodology}} details the design and implementation of our adaptive pulse-level error mitigation algorithm, covering pulse representation, the genetic algorithm framework, fitness evaluation, implementation specifics, and integration with quantum circuits. \textbf{Section~\ref{sec:experimental_setup}} outlines the experimental setup, including benchmark quantum algorithms (Deutsch-Jozsa and Grover), circuit descriptions, noise modeling in simulations, and simulation parameters. \textbf{Section~\ref{sec:results}} presents the experimental results, showcasing fidelity evolution and comparative analyses between optimized and non-optimized pulses for both algorithms. \textbf{Section~\ref{sec:discussion}} discusses the implications of our findings, assessing the strengths and limitations identified in the protocol, and synthesizing overall insights while proposing directions for future research. Finally, \textbf{Section~\ref{sec:conclusion}} concludes by summarizing the key contributions and highlighting the potential impact of our work on advancing quantum error mitigation strategies.

\section{Background and Preliminaries} \label{sec:background}

In this section, we provide the necessary theoretical and technical background behind our method. We begin with an overview of noise modeling in quantum computing, highlighting various types of errors and their mathematical representations. We then discuss techniques for quantum error mitigation (QEM) and contrast these methods with conventional quantum error correction (QEC) schemes. Finally, we introduce QuTiP (Quantum Toolbox in Python), the open-source framework employed for simulating open quantum systems and implementing pulse-level strategies.

\subsection{Noise Modeling in Quantum Computing}
\label{sec:noise-modeling-qc}

Before delving deeper into the mathematical formalisms, we provide a brief roadmap of this subsection by
\begin{itemize}
    \item Reaffirming the significance of noise in NISQ devices.
    \item Presenting the Lindblad and Kraus frameworks, essential tools for describing open quantum system dynamics.
    \item Reviewing standard noise channels (bit-flip, phase-flip, depolarizing, and others) and outline their parameters and typical use cases.
\end{itemize}

This overview will guide the reader from general considerations to the concrete mathematical models of noise typically used in quantum error mitigation and correction.

Quantum computing devices are intrinsically vulnerable to environmental noise and operational imperfections, both of which degrade the fidelity of quantum operations. As discussed by Nielsen and Chuang~\cite{b1}, and highlighted by Preskill in the NISQ era context~\cite{b2}, this susceptibility arises from unavoidable interactions between the quantum system and its environment, as well as practical limitations in our ability to implement qubits, gates and measurements. Understanding and modeling these noise mechanisms across multiple qubit modalities constitute central challenges in developing effective Quantum Error Correction (QEC) and Quantum Error Mitigation (QEM) techniques.

A key theoretical framework used to describe the evolution of open quantum systems under noise and decoherence is the Lindblad master equation. In the Markovian and weak-coupling regime—approximations often valid for contemporary quantum hardware—one can write the time evolution of the density matrix \(\rho(t)\) as~\cite{b13,b20}:

\begin{equation}
\frac{d\rho(t)}{dt} = -i[H, \rho(t)] 
+ \sum_k \left( L_k \rho(t) L_k^\dagger - \tfrac{1}{2}\{L_k^\dagger L_k, \rho(t)\} \right),
\label{eq:lindblad}
\end{equation}

where \(H\) is the system Hamiltonian and \(L_k\) are Lindblad (collapse) operators encoding different decoherence channels. The Lindblad equation provides a continuous-time description of noisy dynamics, capturing processes such as energy relaxation and dephasing.

An equivalent yet flexible viewpoint is offered by the operator-sum or Kraus representation of quantum channels~\cite{b1,b23}. Any completely positive, trace-preserving (CPTP) map \(\mathcal{E}\) acting on \(\rho\) can be expressed as:

\[
\mathcal{E}(\rho) = \sum_\ell E_\ell \rho E_\ell^\dagger, \quad \sum_\ell E_\ell^\dagger E_\ell = I.
\]

Kraus operators \(\{E_\ell\}\) allow one to switch between the Lindblad and operator-sum representations, providing a versatile framework for modeling a wide variety of noise processes. Below, we outline common noise mechanisms and their standard models, following closely Nielsen and Chuang's~\cite{b1} description and related works:

\subsubsection{Decoherence}

\emph{Decoherence} refers to the process by which a quantum system loses coherence due to unwanted interactions with its environment, transitioning from quantum superpositions to classical mixtures~\cite{b19}. This fundamentally limits the time over which arbitrary quantum information can be reliably stored. In the Bloch sphere, decoherence typically contracts the sphere along certain directions, eroding the off-diagonal elements of the density matrix and effectively destroying quantum correlations.

\subsubsection{Relaxation (Amplitude Damping)}

\emph{Amplitude damping} models energy relaxation, such as the decay of an excited state \(|1\rangle\) to the ground state \(|0\rangle\). Characterized experimentally by a timescale \(T_1\), this process is expressed by the Lindblad operator \(L_{\text{relax}} = \sqrt{1/T_1}\,\sigma^-\), where \(\sigma^- = |0\rangle\langle 1|\). In Kraus form~\cite{b1}:

\[
E_0 = \begin{pmatrix}1 & 0 \\ 0 & \sqrt{1-\gamma}\end{pmatrix}, \quad E_1 = \begin{pmatrix}0 & \sqrt{\gamma} \\ 0 & 0\end{pmatrix},
\]

with \(\gamma\) related to the damping probability. Amplitude damping drives states toward \(|0\rangle\), reducing excitation probability and reflecting energy dissipation into the environment.

\subsubsection{Dephasing (Phase Damping)}

\emph{Dephasing}, affects the relative phase between computational basis states without energy exchange. It is characterized experimentally by \(T_2\) and can be modeled by a Lindblad operator \(L_{\text{dephase}} = \sqrt{1/(2T_2)}\,\sigma_z\). In Kraus form~\cite{b1}:
\[
E_0 = \sqrt{1-\lambda}I, \quad E_1 = \sqrt{\lambda}\begin{pmatrix}1 & 0 \\ 0 & 0\end{pmatrix}.
\]

Dephasing destroys off-diagonal elements of \(\rho\), diminishing quantum coherence. \(T_2 \leq 2T_1\) often indicates that phase coherence typically decays as fast as or faster than population relaxation.

\subsubsection{Discrete Pauli Errors (Bit-Flip, Phase-Flip, Bit-Phase-Flip)}

\emph{Discrete Pauli errors} are among the simplest quantum noise models~\cite{b1}. These occur when a Pauli operator \(\sigma_x\), \(\sigma_y\), or \(\sigma_z\) is applied to the state with probability \(p\), leaving the state unchanged with probability \(1-p\). Physically, these can be viewed as sudden, discrete kicks rather than gradual evolutions. Mathematically:

\[
\mathcal{E}(\rho) = (1-p)\rho + p(\sigma\rho\sigma),
\]

with \(\sigma \in \{\sigma_x, \sigma_y, \sigma_z\}\). In Kraus form:

\[
E_0 = \sqrt{1-p}\,I, \quad E_1 = \sqrt{p}\,\sigma.
\]

Specific channels include:

\begin{itemize}
    \item \textbf{Bit-Flip:} Applies \(\sigma_x\) with probability \(p_{\text{bit-flip}}\), flipping \(|0\rangle \leftrightarrow |1\rangle\).
    \item \textbf{Phase-Flip:} Applies \(\sigma_z\) with probability \(p_{\text{phase-flip}}\), introducing a relative phase between \(|0\rangle\) and \(|1\rangle\).
    \item \textbf{Bit-Phase-Flip:} Applies \(\sigma_y\) with probability \(p_{\text{bit-phase-flip}}\), simultaneously flipping the bit value and its relative phase.
\end{itemize}

The discrete channels above constitute composable building blocks useful to model more complex noise sources. They are central to the theory of quantum error correction, where correcting these basic errors implies the ability to correct arbitrary single-qubit errors due to the Pauli basis decomposition. In our case, we make use of the Kraus representation extensively in the sections below.

\subsubsection{Depolarizing Error}

The \emph{depolarizing channel} is a symmetric noise model often used for benchmarking~\cite{b1}. With probability \(p\), it replaces \(\rho\) by the maximally mixed state \(I/2\), and with probability \(1-p\) leaves it unchanged:

\[
\mathcal{E}_{\text{depol}}(\rho) = (1-p)\rho + \frac{p}{3}(\sigma_x\rho\sigma_x + \sigma_y\rho\sigma_y + \sigma_z\rho\sigma_z).
\]

Geometrically, the depolarizing channel uniformly contracts the Bloch sphere toward its center, modeling a highly isotropic form of noise.

\subsubsection{Parameters in Simulations}

In practical simulations, incorporating noise involves specifying parameters and channels that reflect the target device or scenario:
\begin{itemize}
    \item \(N\): Number of qubits.
    \item \(T_1\): Relaxation time (amplitude damping).
    \item \(T_2\): Dephasing time.
    \item Probabilities \(p_{\text{bit-flip}}, p_{\text{phase-flip}}, p_{\text{depol}}\), etc., governing discrete error channels.
\end{itemize}

By selecting the appropriate either Lindblad or Kraus operators through experiments and device characterizations, one can realistically simulate noisy quantum circuits. Doing so enables the evaluation of quantum algorithms in the presence of noise and guides the development of QEC and QEM strategies essential for achieving robust quantum computations~\cite{b1,b2}. For instance, consider the application of an arbitrary single-qubit gate subjected to a depolarizing channel with probability $p=0.01$. The corresponding noise channel uniformly shrinks the Bloch sphere, slightly reducing the fidelity of any intended gate. In the context of the work presented in this manuscript, understanding the interplay between noise sources and pulse-level control is critical. As we will describe in Section~\ref{sec:methodology}, our adaptive genetic algorithm optimizes pulse parameters in real-time to mitigate the impact of noise, including scenarios in which a small depolarizing probability can accumulate and degrade performance as the case described above.

Since NISQ devices lack the resources for large-scale quantum error correction, employing Lindblad- and Kraus-based models at the pulse level offers a pragmatic alternative. By working directly with these representations in a well-supported simulation framework such as QuTiP, we can dynamically shape pulses to evolving noise conditions. This integration of theoretical noise modeling with pulse-level adaptive optimization sets the stage for the techniques introduced in Section~\ref{sec:methodology}.

\subsection{State of the Art in Quantum Error Modeling}

Contemporary research in quantum computing has produced increasingly refined methods to characterize and simulate noise at different levels of abstraction. Early works primarily considered simplified error channels, such as depolarizing or amplitude damping processes, introduced solely at the gate level \cite{b1,b2,b3,b20,b24,b25}. Although these simplified representations were intended to capture essential decoherence mechanisms, they often neglect the intricate interplay between noise and unitary evolution that takes place during the execution of gates. As highlighted by Di Bartolomeo \textit{et al.}~\cite{Di_Bartolomeo_2023}, representing noise purely as separate stochastic operations interspersed between ideal gates may fail to fully reflect the underlying Markovian (and more generally, open-system) dynamics that real quantum hardware undergoes.

Recent frameworks and quantum software toolkits have expanded their capabilities to incorporate more realistic noise models. Standard tools like \textit{Qiskit} \cite{qiskit}, \textit{Cirq} \cite{cirq2020}, and \textit{Amazon Braket} \cite{amazonBraket} provide built-in functionality to insert Kraus operators or Lindblad-like channels into simulated circuits, enabling a more faithful rendition of device physics. These implementations, however, commonly maintain a strict separation between the intended unitary operations and stochastic error processes \cite{harper2020efficient, strikis2021learning}. Although effective at capturing state-dependent decay channels and certain adaptive calibration protocols, these methods still presuppose a modular composition of errors that does not inherently account for the continuous-time influence of noise during gate pulse execution.

Additional refinements to noise modeling emphasize adaptive calibration and device characterization. For instance, quantum noise tomography \cite{harper2020efficient} and learning-based noise inference \cite{strikis2021learning} seek to extract more detailed noise parameters, potentially incorporating correlations and non-Markovian features. Yet, these techniques typically remain attached to frameworks that insert noise at discrete points in the circuit.

As discussed by Di Bartolomeo \textit{et al.}~\cite{Di_Bartolomeo_2023}, one can incorporate noise directly into gate definitions by modifying the time evolution that generates each gate. Instead of treating noise as extraneous channels appended before or after an ideal unitary, their formalism suggests building 'noisy gates' that arise naturally from solving the time-dependent Lindblad equation. This perspective ensures that the resultant operations automatically encode the interplay between unitary control and environmental coupling; this paradigm potentially outperforms traditional gate-plus-channel schemes, particularly in regimes where gate durations are not negligible compared to decoherence timescales. While we did not pursue this path in the work reported here due to time constraints, it is a natural extension we will address in future research.

The current frontier in quantum computing aims to go beyond traditional static gate-level noise toward noise modeling directly at the pulse level. The resulting paradigm, further informed by advances in tomography, machine learning, and adaptive calibration, paves the way for new strategies in how quantum gates are implemented. Embedding noise-aware techniques into gate design allows experimentalists and theorists to advance toward simulations that more faithfully represent the behavior of NISQ devices. All these strategies ultimately contribute to increase the realism of quantum simulations, reshape strategies for error mitigation, and ultimately enable quantum algorithms with greater physical fidelity and practicality.

\subsection{Quantum Error Mitigation and Correction Techniques}

Quantum Error Correction (QEC) seeks to protect quantum information from errors by redundantly encoding logical qubits into multiple physical qubits. Classical syndrome measurements lead to detecting and correcting certain types of errors without collapsing the encoded quantum state. Reliable quantum computation is possible whenever error rates are below a known thresholds~\cite{b4}. Albeit successful, QEC implementations are resource-intensive and require significant overhead in terms of qubit count and error rates that are still challenging to achieve on current hardware.

In contrast, Quantum Error Mitigation (QEM) aims to reduce the impact of noise without fully correcting it. Instead of achieving fault-tolerance, QEM employs post-processing techniques, calibration methods, and noise extrapolation schemes to partially compensate for errors. For instance, zero-noise extrapolation linearly projects results from multiple runs with artificially amplified noise back to a zero-noise scenario, while probabilistic error cancellation statistically eliminates certain noise contributions~\cite{b9}. Although QEM does not produce a fully error-corrected logical qubit, it can significantly improve the fidelity of quantum algorithms on NISQ devices, increasing the feasibility of certain algorithms under practical conditions today~\cite{b11,b12}. Nevertheless, as established by Takagi \textit{et al.}~\cite{Takagi_2022}, QEM methods are inherently limited, particularly as circuit depth increases, due to fundamental constraints on the achievable error mitigation without exponential resource overhead. This underscores the complementary nature of QEM and QEC in advancing quantum computing, where QEM offers immediate fidelity improvements within the constraints of NISQ devices, and QEC provides a scalable path toward fault-tolerant quantum computation.

Our method builds on QEM strategies by focusing directly on the control fields that translate quantum dynamics into operations (gates) directly at the pulse level. This allows us to adapt to noise profiles in real-time, addressing variations in noise profiles without altering circuit architecture or increasing qubit overhead. Our adaptive control method complements existing QEM and QEC techniques.

\subsection{Introduction to QuTiP} \label{sec:qutip}

To implement and evaluate our adaptive pulse-level quantum error mitigation technique, we rely on \emph{Quantum Toolbox in Python} (\texttt{QuTiP})~\cite{b10}, an open-source  framework for simulating the dynamics of open quantum systems. \texttt{QuTiP} accommodates various tasks that include noise modeling, time-dependent Hamiltonian evolution, and real-time tracking of quantum states under user-defined control parameters.

\subsubsection{Abstract model}

A principal component of \texttt{QuTiP} is the \texttt{Qobj} class, which unifies quantum states, operators, and superoperators into a coherent interface~\cite{b9,b10}. Being an object-oriented framework enables sophisticated manipulations (e.g., tensor products, 
partial traces, matrix exponentials, and spectral decompositions) using intuitive high-level syntax. For instance, composite systems of multiple qubits can be constructed seamlessly by tensoring single-qubit states or operators.

\subsubsection{Open-System Dynamics}

Another hallmark feature of \texttt{QuTiP} is its capacity to model open-system dynamics via master equation solvers. The Lindbladian introduced in Section~\ref{sec:background} is usually numerically integrated using the \texttt{mesolve} function by considering the alternative form

\begin{equation}
  \frac{d\rho(t)}{dt} 
  = -\,i\bigl[H(t),\,\rho(t)\bigr] 
    + \sum_{j}\Bigl( 
    C_j\,\rho(t)\,C_j^\dagger 
    - \tfrac{1}{2}\bigl\{C_j^\dagger C_j, \rho(t)\bigr\}
    \Bigr),
\end{equation}

where \(H(t)\) is a (potentially time-dependent) Hamiltonian and \(\{C_j\}\) are collapse operators capturing decoherence. In this work, we make use of the \texttt{mesolve} solver, which provides an ensemble-averaged picture of open-system evolution. Meanwhile, \texttt{mcsolve} offers a more granular alternative that unravels the master equation into individual quantum trajectories, capturing stochastic fluctuations that may be obscured in ensemble-averaged treatments. Although \texttt{mcsolve} is especially pertinent when one wishes to model pulse-level noise events on a single-run basis, we chose \texttt{mesolve} in our study to focus on the averaged dynamics and simplify computational overhead. 
 \cite{lambert2024qutip5quantumtoolbox}

\subsubsection{Integration and Performance}

\texttt{QuTiP} integrates efficiently with Python libraries such as \texttt{NumPy},  \texttt{SciPy}, and \texttt{Matplotlib}, allowing seamless transitions from problem formulation to numerical simulation and data visualization.

\vspace{0.8em}

\subsection{Introduction to QuTiP-QIP} \label{sec:qutip-qip}

Building upon the functionalities of \texttt{QuTiP}, the \texttt{qutip-qip} module~\cite{b9} extends the support to the simulation of pulse-level circuits. The package is intended to model quantum hardware at the time-evolution level, treating gates as time-sequenced control pulses while incorporating noise processes inherent to realistic devices.

\subsubsection{Processor - Pulse-Level Simulations}

Central to \texttt{qutip-qip} is the notion of a \texttt{Processor}, which encapsulates both drift and control Hamiltonians:

\begin{equation}
  H(t) \;=\; H_{d} \;+\;\sum_{j}\;u_{j}(t)\,H_{j},
\end{equation}

where \(H_d\) represents the intrinsic system Hamiltonian, and \(\{H_j\}\) are 
control Hamiltonians modulated by time-dependent coefficients \(u_j(t)\). This structure enables direct simulation of pulse dynamics, including imperfections and decoherence.

\subsubsection{OptPulseProcessor and Key Parameters}

\texttt{OptPulseProcessor}, 
which applies optimal control methods to refine \(u_j(t)\) for desired gate operations specified as pulse sequences, even in the presence of noise. Two parameters are of special interest:

\begin{itemize}
  \item \textbf{Evolution Time} (\texttt{evo\_time}): The total duration over which the quantum state evolves under \(H(t)\). This defines how long the system interacts with both drift and control fields.

  \item \textbf{Number of Time Slices} (\texttt{num\_tslots}): the number of segments in which time evolution is to be discretized for the fine-tuning of \(u_j(t)\) at each step. A larger number of segments improves the resolution of pulse shaping at the expense of increased computational resources.
\end{itemize}

Mathematically, the unitary evolution over an interval \(\Delta t\) satisfies

\begin{equation}
  U(\Delta t) 
  \;=\; \exp\!\Bigl(\,
  i \,\Delta t\;\bigl[\,H_{d}\;+\;\sum_{j}\,u_{j}\,H_{j}\bigr] 
  \Bigr).
\end{equation}

Adjusting \(u_j\) across \texttt{num\_tslots} within a fixed \texttt{evo\_time}, results in operations which can approximate target gates with high fidelity.

\subsubsection{Simulation Flow}

A QuTiP-based simulation consists of three main steps performed in sequence:

\begin{enumerate}
  \item \textbf{Circuit Compilation}: Quantum gates are mapped onto Hamiltonian-based pulses that account for the hardware’s drift.
  \item \textbf{Noise Integration}: Decoherence models (e.g., collapse operators) 
    or error probabilities are introduced, leveraging \texttt{QuTiP}'s solvers.
  \item \textbf{Control Optimization and Evolution}: 
    The \texttt{OptPulseProcessor} optimizes \(\{u_j(t)\}\) over time slices. 
    \texttt{mesolve} then integrates the master equation over the specified time range to obtain the final state or measurement statistics.
\end{enumerate}

This modular framework includes several \texttt{Processor} subclasses for specific physical systems:
\begin{itemize}
  \item \textbf{SpinChain}: Predefined \(\sigma_x, \sigma_z,\) and two-qubit 
    interactions suited for spin-chain devices.
  \item \textbf{DispersiveCavityQED}: Multi-level cavity models with qubit-cavity coupling. 
  \item \textbf{OptPulseProcessor}: Optimally controlled quantum processor intended to refine pulses, aiming for robust gate fidelity amid noise typical of NISQ scenarios.
\end{itemize}

Altogether, \texttt{qutip-qip} bridges theoretical circuit abstractions with 
realistic device physics, enabling pulse-level simulations that align with 
the constraints and error models pertaining to current quantum hardware.

\section{Methodology} \label{sec:methodology}

In this section, we provide a detailed description of the design and implementation of our adaptive pulse-level error mitigation algorithm. The corresponding code has been made available in a GitHub repository under the open-source MIT License as indicated in \textit{Code and Data Availability}.

\subsection{Pulse Representation and Control}

A pulse is a time series of complex-valued amplitudes with a maximum unit norm, represented as \([d_0, \dots, d_{n-1}]\). Each \(d_j\), where \(j \in \{0, \dots, n-1\}\), is called a sample. Each system specifies a cycle time \(dt\), which is the finest time resolution available in the pulse coprocessor. This cycle time is typically defined by the sampling rate of the coprocessor's waveform generators. Each sample in a pulse is issued over one cycle, corresponding to a single time step. All pulse durations and time steps are defined and discretized with respect to \(dt\).

The ideal output signal has an amplitude given by:

\begin{equation}
D_j = \text{Re}\left[ e^{i(2\pi f_j dt + \phi_j)} \right]
\label{eq:pulse_amplitude}
\end{equation}

at time \(f_j dt\), where \(f_j\) is a modulation frequency and \(\phi_j\) is a phase. Pulse samples describe only the envelope of the signal produced, which is then mixed in hardware with a carrier signal defined by its frequency and phase \cite{b14}.

\subsection{Adaptive Pulse-Level Error Mitigation Algorithm} \label{sec:algorithm}

We present below an overview of the Adaptive Pulse-Level Error Mitigation Algorithm, which leverages an enhanced Genetic Algorithm framework to dynamically optimize pulse parameters. The algorithm operates over multiple generations, evolving a population of candidate solutions through selection, crossover, mutation, and other genetic operations. Each component of the algorithm is detailed in the Appendix (see Appendix~\ref{appendix:algorithm}).

\begin{algorithm}
\caption{Adaptive Pulse-Level Error Mitigation Algorithm (Overview)}
\label{alg:main_algorithm}
\small
\SetAlgoLined
\SetKwInOut{KwData}{Input}
\SetKwInOut{KwResult}{Output}
\KwData{
    Population size $N$, 
    Number of generations $G$, 
    Initial mutation probability $p_{\text{mut}}$, 
    Initial crossover probability $p_{\text{cross}}$, 
    Feedback threshold $\delta$, 
    Feedback interval $I$, 
    Early stopping rounds $R$, 
    Diversity threshold $\theta$, 
    Diversity action (`mutate' or `replace')
}
\KwResult{Optimized pulse parameters maximizing fidelity}

Initialize population $P_0$ \tcp*{See Algorithm~\ref{alg:initialization}}\;
\For{$g \leftarrow 1$ \KwTo $G$}{
    Select parents using Tournament Selection \tcp*{See Algorithm~\ref{alg:selection}}\;
    Apply Crossover \tcp*{See Algorithm~\ref{alg:crossover}}\;
    Apply Mutation \tcp*{See Algorithm~\ref{alg:mutation}}\;
    Evaluate Fitness \tcp*{See Algorithm~\ref{alg:fitness_evaluation}}\;
    Replace Population \tcp*{See Algorithm~\ref{alg:replacement}}\;
    Apply Elitism \tcp*{See Algorithm~\ref{alg:elitism}}\;
    Control Diversity \tcp*{See Algorithm~\ref{alg:diversity_control}}\;
    Adjust Parameters \tcp*{See Algorithm~\ref{alg:feedback_mechanism}}\;
    Check Early Stopping \tcp*{See Algorithm~\ref{alg:early_stopping}}\;
}
\Return{Best individual (i.e., highest fidelity)}\;
\end{algorithm}

Our primary contribution is the development of an \textbf{Adaptive Pulse-Level Error Mitigation Algorithm} based on an enhanced Genetic Algorithm (GA) framework. This algorithm dynamically optimizes pulse parameters to mitigate quantum errors induced by noise and decoherence in quantum circuits. Key features of our algorithm include:

\begin{itemize}
    \item \textbf{Feedback-Based Mutation and Crossover Adjustment:} We incorporate a dynamic feedback mechanism to adaptively adjust mutation and crossover probabilities, $p_{\text{mut}}$ and $p_{\text{cross}}$, based on observed improvement in the average fitness of the population over successive generations \cite{b8}. Specifically, after every $I$ generations, we compute the change in average fitness $\Delta \overline{F} = \overline{F}_{g} - \overline{F}_{g-I}$. If $\Delta \overline{F}$ falls below a predefined improvement threshold $\delta$, indicating stagnation in the search process, we increase $p_{\text{mut}}$ and $p_{\text{cross}}$ by a small increment $\Delta p$ to promote diversity and exploration of new regions in the solution space. Conversely, if significant improvement is observed ($\Delta \overline{F} \geq \delta$), we decrease $p_{\text{mut}}$ and $p_{\text{cross}}$ to fine-tune the search around promising areas, emphasizing exploitation. Mathematically, the adjustment rule can be expressed as:
    
    \[
    p_{\text{mut}}^{(g+1)} = \begin{cases}
    p_{\text{mut}}^{(g)} + \Delta p, & \text{if } \Delta \overline{F} < \delta \\
    p_{\text{mut}}^{(g)} - \Delta p, & \text{otherwise}
    \end{cases}
    \]
    
    \[
    p_{\text{cross}}^{(g+1)} = \begin{cases}
    p_{\text{cross}}^{(g)} + \Delta p, & \text{if } \Delta \overline{F} < \delta \\
    p_{\text{cross}}^{(g)} - \Delta p, & \text{otherwise}
    \end{cases}
    \]
    
Our adaptive mechanism ensures a balanced search strategy that dynamically responds to the pattern of variation of improvement during the optimization progress. This prevents the algorithm from being trapped by local optima and enhances convergence towards the global optimum.

    \item \textbf{Diversity Control}
    To prevent premature convergence and maintain genetic diversity within the population, we implement diversity control strategies that monitor the genetic variance of the population \cite{b18}. We calculate the diversity $D$ using the \textbf{average Mahalanobis distance} between all pairs of individuals in the parameter space.  The choice of the average Mahalanobis distance is motivated by its ability to account for correlations between different parameters and provide a scale-invariant measure of diversity. Unlike simpler metrics such as Euclidean distance, the Mahalanobis distance considers the covariance structure of the population, allowing for a more nuanced assessment of genetic variance.The Mahalanobis distance between two individuals $\mathbf{x}_i$ and $\mathbf{x}_j$ is defined as:
    
    \[
    D_M(\mathbf{x}_i, \mathbf{x}_j) = \sqrt{(\mathbf{x}_i - \mathbf{x}_j)^T \Sigma^{-1} (\mathbf{x}_i - \mathbf{x}_j)}
    \]
    
    where $\Sigma$ is the covariance matrix of the variables in the population. Diversity $D$ is then the average of Mahalanobis distances across all unique pairs of individuals:
    
    \[
    D = \frac{2}{N(N-1)} \sum_{i=1}^{N-1} \sum_{j=i+1}^{N} D_M(\mathbf{x}_i, \mathbf{x}_j)
    \]
    
    where $N$ is the population size and $D_M(\mathbf{x}_i, \mathbf{x}_j)$ is the Mahalanobis distance between individuals $i$ and $j$. If $D$ falls below a predefined threshold $\theta$, indicating that individuals are becoming too similar, we apply one of the following actions:
    
    \begin{itemize}
        \item \textbf{Higher Variance Mutation:} Increase mutation rate to reintroduce genetic variability.
        \item \textbf{Individual Replacement:} Replace a percentage of the population with new random individuals.
    \end{itemize}
    
    This strategy ensures that the algorithm continues to effectively explore the solution space, increasing the likelihood of discovering high-quality solutions that might otherwise be overlooked.

    \item \textbf{Elitism:} The algorithm employs an elitist strategy by retaining the top performing individual(s) from each generation and directly transferring them to the next generation without alteration. This ensures that the best-found solutions are preserved throughout the evolutionary process, preventing the loss of valuable genetic information due to stochastic operations like mutation and crossover. By maintaining an elite set of solutions, we ensure that the algorithm's performance does not degrade over time and that high-fidelity pulse parameters are continuously refined and propagated.
    
    \item \textbf{Early Stopping Criterion:} To enhance computational efficiency and prevent unnecessary evaluations, we implement an early stopping mechanism that monitors the improvement of the best fitness value over consecutive generations. If no significant improvement (exceeding a minimal threshold $\epsilon$) is observed in the best fitness after $R$ consecutive generations, we conclude that the algorithm has likely converged to an optimal or near-optimal solution and terminate the optimization process. Formally, if
    
    \[
    F_{\text{best}}^{(g)} - F_{\text{best}}^{(g - R)} < \epsilon
    \]
    
    then the algorithm stops. This criterion helps save computational resources, especially important given the high computational cost associated with simulating quantum systems and evaluating the fitness function.

    \item
 
    \textbf{Parallelization:}  Recognizing the computational 
    intensity of simulating quantum circuits for fitness evaluations, we adopt a dynamic work-stealing strategy via the \texttt{SCOOP} framework, adopting the framework introduced by Hold-Geoffroy \textit{et al.}~\cite{scoop}. Under this paradigm, each fitness evaluation constitutes a self-contained task with no inter-task communication. SCOOP relies on a \textbf{broker--worker} \cite{patternorientedsa} design: \begin{itemize} \item \textbf{Broker process:} Manages global scheduling by receiving and redistributing tasks. It connects to workers through a router pattern, holding a central view of the workload to ensure dynamic load balancing. In addition, a publisher socket handles one-to-many broadcast messages, such as requests to terminate or directives to share constants. \item \textbf{Worker processes:} Execute tasks in a decentralized manner. Each worker subscribes to the broker’s messages and cooperates with other workers by forming a peer-to-peer network. When a worker finishes existing tasks and detects surplus demand, it requests more tasks from the broker, effectively “stealing” work. This adaptively levels out the load across all available resources without requiring explicit “fork” or “join” calls. \end{itemize} This architecture enables SCOOP to distribute hierarchical or nested tasks automatically --each fitness evaluation can spawn subtasks, which in turn can spawn further subtasks—without the user explicitly managing inter-process communication. By shifting scheduling and data routing to the broker–worker pattern, the framework seamlessly scales to multi-core or multi-node deployments while minimizing idle CPU time.
\end{itemize}

\subsubsection{Fitness Function}

The fitness function is defined as the quantum state fidelity between the final state \(\rho_{\text{final}}\) obtained after applying the pulse sequence and the target state \(\rho_{\text{target}}\). For arbitrary density matrices \(\rho_{\text{final}}\) and \(\rho_{\text{target}}\), the fidelity is given by the Uhlmann formula

\begin{equation}
F(\rho_{\text{final}}, \rho_{\text{target}}) 
= \left( \mathrm{Tr}\sqrt{\sqrt{\rho_{\text{final}}}\,\rho_{\text{target}}\,\sqrt{\rho_{\text{final}}}} \right)^{2}.
\label{eq:fitness_function}
\end{equation}

In the special case where both states are pure, \(\rho_{\text{final}} = |\psi_{\text{final}}\rangle \langle \psi_{\text{final}}|\) and \(\rho_{\text{target}} = |\psi_{\text{target}}\rangle \langle \psi_{\text{target}}|\), the fidelity reduces to

\[
F(\rho_{\text{final}}, \rho_{\text{target}}) 
= |\langle \psi_{\text{target}} | \psi_{\text{final}} \rangle|^{2}.
\]

\subsubsection{Implementation Details}

We implemented the algorithm using Python, leveraging the DEAP (Distributed Evolutionary Algorithms in Python) library for the GA framework \cite{b15}. Integration with QuTiP allowed us to simulate quantum circuits and accurately evaluate the fitness of individuals. Key implementation details in Python follow.

\begin{itemize}
    \item \textbf{Parallelization:} We utilized the \texttt{SCOOP} framework to efficiently parallelize the evaluation of individuals across multiple CPU cores. By leveraging SCOOP's broker–worker design, tasks were seamlessly distributed among available resources, enabling concurrent processing without the need for explicit inter-process communication management.
    \item \textbf{Random Number Generation:} To ensure high-quality randomness, we used random generator provided by \texttt{numpy} alongside the  \texttt{secrets} module.
    \item \textbf{Custom Operators:} We defined custom crossover and mutation operators tailored to the structure of the pulse parameters.
    \item \textbf{Logging and Monitoring:} The algorithm logs performance metrics, such as average fitness, standard deviation, and maximum fitness, for each generation, facilitating analysis and visualization.
\end{itemize}

\subsubsection{Algorithm Workflow}
\label{sec:algorithm_workflow}

\begin{figure}[h]
    \centering
    \includegraphics[width=\linewidth]{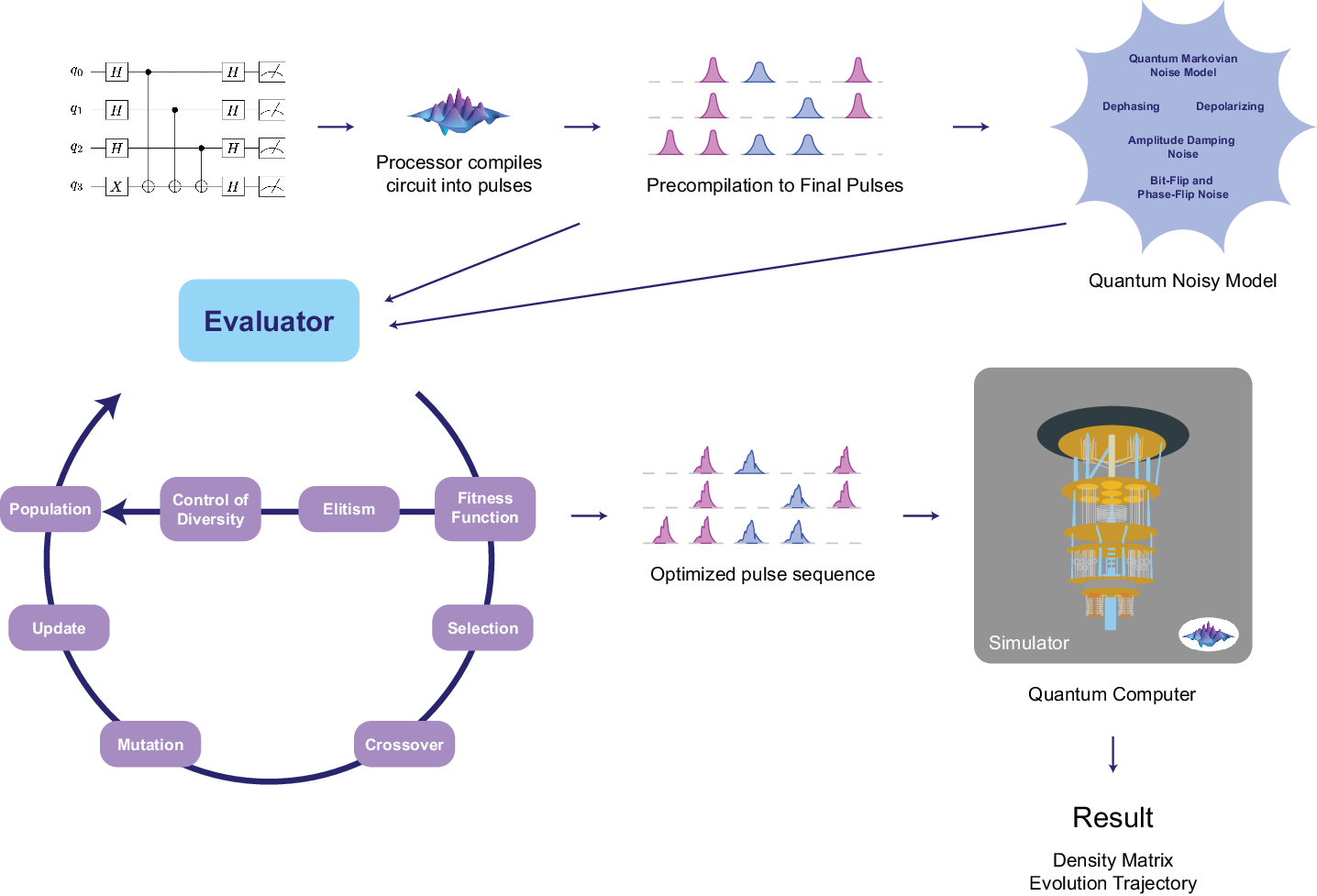}
    \caption{Workflow of the Adaptive Pulse-Level Error Mitigation Algorithm.}
    \label{fig:algorithm_flowchart}
\end{figure}

The intended workflow is illustrated in Figure~\ref{fig:algorithm_flowchart}, which begins with a quantum circuit. The \texttt{qutip-qip} ~\cite{b9} framework then translates this circuit into precompiled pulse sequences. At this stage, we inject quantum Markovian noise (described in detail in Section~\ref{sec:noise-modeling-qc}), which includes various noise models --i.e., dephasing, depolarization, and amplitude damping. Next, we use an evaluator to measure the quantum fidelity \cite{b22} of the system under these noisy conditions. This fidelity serves as the objective for our Adaptive Genetic Algorithm (AGA), which iteratively optimizes the pulse parameters to maximize the fidelity. The result is an optimized pulse sequence that reflects the refinements introduced by the AGA. Finally, these optimized pulses are fed into a quantum simulator, which computes the resulting density matrix and evolution trajectory. This output provides a detailed characterization of the quantum system's performance under the optimized conditions.

\subsection{Integration with Quantum Circuits}

We applied the genetic optimization algorithm to refine pulse parameters for two established quantum algorithms, namely Grover's search \cite{b17} and  Deutsch-Jozsa \cite{b16}. This refinement specifically targets the evolution time (\texttt{evo\_time}) and the number of time slices (\texttt{num\_tslots}) associated with pulse sequences implementing each quantum gate, as introduced in Section~\ref{sec:qutip-qip}. By meticulously adjusting these temporal parameters, the algorithm enhances the fidelity of gate operations in noisy environments, thereby improving the overall performance and robustness of the implemented quantum circuits.

\FloatBarrier

\section{Experimental Setup} \label{sec:experimental_setup}

In this section, we describe the experimental setup used to evaluate our algorithm. To ensure scalability and efficiency, all experiments were conducted on a MacBook Pro equipped with Apple’s M3 chip, which offers integrated CPU, GPU, and Neural Engine capabilities on a single System on Chip (SoC). Table~\ref{tab:vm_specs} summarizes the main hardware specifications, demonstrating sufficient computational resources to handle pulse-level quantum simulations and the parallelized fitness evaluations in our genetic algorithm framework.

\begin{table}[h!]
\centering
\caption{MacBook Pro M3 Hardware Specifications}
\label{tab:vm_specs}
\begin{tabular}{l l}
\hline
\textbf{Component} & \textbf{Specification} \\
\hline
System on Chip (SoC)       & Apple M3 \\
CPU Cores                  & 8-core (4 performance + 4 efficiency) \\
GPU Cores                  & 10-core GPU \\
Neural Engine              & 16-core \\
Memory Bandwidth           & 100 GB/s \\
RAM                        & 24 GB \\
\hline
\end{tabular}
\end{table}

\subsection{Benchmark Quantum Algorithms}

Well-known quantum algorithms serve as benchmarks for evaluating quantum hardware performance, gate fidelity, and coherence times. Implementing these algorithms at the pulse level on actual quantum devices allows researchers to gather valuable metrics on the scalability, robustness, and accuracy of their experimental setups. Two particularly instructive examples are the Deutsch-Jozsa algorithm, which provides an exponential speedup when determining whether a Boolean function is constant or balanced (\cite{b16}), and Grover’s algorithm, which offers a theoretical quadratic speedup for unstructured database search (\cite{b17}).

A key measure for benchmarking is the fidelity (\cite{b1}) between quantum states. Fidelity quantifies how close two quantum states are and serves as a useful tool for assessing the quality of the implemented quantum operations. The fidelity \(F(\rho,\sigma)\) between two density matrices \(\rho\) and \(\sigma\) is defined as

\[
F(\rho,\sigma) \equiv \left(\mathrm{tr}\sqrt{\rho^{1/2}\sigma\rho^{1/2}}\right)^2.
\]

It is symmetric, bounded between 0 and 1, invariant under unitary transformations, and reduces to the overlap between states when one of them is pure. High fidelity indicates that the prepared or evolved state closely matches the ideal target state, making fidelity a natural figure of merit for comparing expected algorithmic outcomes with those observed in experiments.

\subsection{Deutsch-Jozsa Algorithm}

The Deutsch-Jozsa algorithm (\cite{b16}) determines whether a given Boolean function \( f: \{0,1\}^{n} \to \{0,1\} \) is constant or balanced with a single evaluation. With four qubits, three are used as inputs and one as an ancilla. The algorithm produces a clear and deterministic expected result, which makes it an excellent initial benchmark. If implemented correctly, measuring the final state after the algorithm indicates whether the function is constant (e.g., ideally measuring \(|0000\rangle\)) or balanced. The fidelity between the experimentally obtained final state and the ideal theoretical outcome quantifies the quality of the hardware and pulse-level control as simulated under various noise regimes.

\begin{figure}[h!]
\centering
\[
\Qcircuit @C=1em @R=1.2em {
    \lstick{\ket{q_0}} 
        & \qw
        & \qw 
        & \gate{H} 
        & \qw 
        & \ctrl{3}
        & \qw 
        & \qw 
        & \gate{H}
        & \meter{} \\
    \lstick{\ket{q_1}} 
        & \qw
        & \qw 
        & \gate{H} 
        & \qw 
        & \qw      
        & \ctrl{2}
        & \qw      
        & \gate{H}
        & \meter{}
         \\
    \lstick{\ket{q_2}} 
        & \qw
        & \qw 
        & \gate{H} 
        & \qw 
        & \qw      
        & \qw      
        & \ctrl{1}
        & \gate{H}
        & \meter{} \\
    \lstick{\ket{q_3}} 
        & \gate{X}
        & \qw 
        & \gate{H}
        & \qw 
        & \targ    
        & \targ
        & \targ
        & \qw
        & \meter{}
}
\]

\caption{Deutsch--Jozsa circuit applied to 4 qubits. The Hadamard gate on \(q_3\) is placed in a separate ancilla from the controlled operations.}
\label{fig:dj_separate_h}
\end{figure}
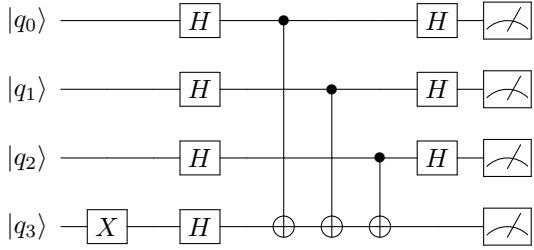

\subsubsection{Circuit Description}

\begin{enumerate}
    \item \textbf{Initialization:} Start with the qubit register in the \(\lvert 0000\rangle\) state.

    \item \textbf{Ancilla Preparation:} 
    Apply an \(X\) gate to qubit~3 to obtain the state \(\lvert 0001\rangle\).

    \item \textbf{Superposition:}
    Apply a uniform superposition $H^{\otimes4}$ to the qubit register.

    \item \textbf{Oracle:}
    Apply an oracle unitary gate $U_f$ implementing a total function $f$. In the example here, $f = XOR$.

    \item \textbf{Superposition:}
    Apply a second uniform superposition $H^{\otimes4}$ to the qubit register after the oracle computation.

    \item \textbf{Measurement:} 
    Measure the qubit register. An \(\lvert 0000\rangle\) outcome
    indicates a constant function, whereas other results indicate a balanced 
    function.
\end{enumerate}

\subsection{Grover's Search Algorithm}

Grover’s algorithm (\cite{b17}) offers a theoretical quadratic speedup for searching an unstructured database. With four qubits, the database size is \(N=16\). The algorithm starts by producing a uniform superposition over all states, applies an oracle that marks a target state \(|1111\rangle\), and then uses the diffusion operator to amplify the amplitude of the marked state. After a suitable number of iterations, measuring the qubits yields the target state with high probability. The fidelity between the experimentally obtained final state distribution and the ideal one reflects how well multiple layers of gates—implemented through pulses—perform in practice.

\begin{figure}[h!]
\centering
\scalebox{0.85}{ 
\Qcircuit @C=0.7em @R=1em {
    \lstick{\ket{q_0}}
        & \gate{H}
        & \gate{X}
        & \ctrl{1}
        & \qw
        & \qw
        & \gate{X}
        & \gate{H}
        & \gate{X}
        & \ctrl{1}
        & \qw
        & \qw
        & \gate{X}
        & \gate{H}
        & \meter{}
        \\
    \lstick{\ket{q_1}}
        & \gate{H}
        & \gate{X}
        & \targ
        & \ctrl{1}
        & \qw
        & \gate{X}
        & \gate{H}
        & \gate{X}
        & \targ
        & \ctrl{1}
        & \qw
        & \gate{X}
        & \gate{H}
        & \meter{}
        \\
    \lstick{\ket{q_2}}
        & \gate{H}
        & \gate{X}
        & \qw
        & \targ
        & \ctrl{1}
        & \gate{X}
        & \gate{H}
        & \gate{X}
        & \qw
        & \targ
        & \ctrl{1}
        & \gate{X}
        & \gate{H}
        & \meter{}
        \\
    \lstick{\ket{q_3}}
        & \gate{H}
        & \gate{X}
        & \qw
        & \qw
        & \targ
        & \gate{X}
        & \gate{H}
        & \gate{X}
        & \qw
        & \qw
        & \targ
        & \gate{X}
        & \gate{H}
        & \meter{}
}
}
\caption{Condensed 4-qubit Grover circuit marking \(\lvert 1111\rangle\). 
The oracle (columns 2--6) adds a phase flip on \(\lvert 1111\rangle\), 
while the diffusion operator (columns 7--13) amplifies that marked state.}
\label{fig:grover4_qubits_condensed}
\end{figure}
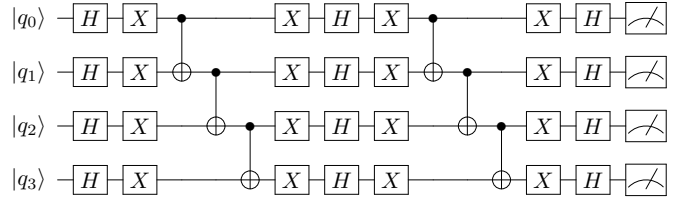

\subsubsection{Circuit Description}

\begin{enumerate}
    \item \textbf{Initialization:}
    Start with the qubit register in the \(\lvert 0000\rangle\) state.

    \item \textbf{Superposition:}
    Apply a uniform superposition $H^{\otimes4}$ to the qubit register.

    \item \textbf{Oracle $U_f$:}
    \begin{enumerate}
        \item \textbf{Preparation:} Apply $X^{\otimes4}$ to the qubit register.
        \item \textbf{Phase Flip (target \(\lvert 1111\rangle\)):} Use three consecutive \(\mathrm{CNOT}\) gates, 
        with controls \((q_i \to q_{i+1})\), $i = 0,1,2$ to add a \(-1\) phase to the prior state.
        \item \textbf{Reversion:} Apply $X^{\otimes4}$ again to restore 
        all non-marked states to their original form and add a global phase of \(-1\) to the target state.
    \end{enumerate}

    \item \textbf{Diffusion Operator:}
    \begin{enumerate}
        \item \textbf{Superposition:} Apply a uniform superposition $H^{\otimes4}$ to the qubit register.
        \item \textbf{Oracle:} Apply the oracle $U_f$ constructed prior.
        \item \textbf{Superposition:} Apply a uniform superposition $H^{\otimes4}$ to the qubit register to reinforce the amplitude of the marked target state.
    \end{enumerate}

    \item \textbf{Iteration:}
    Apply the combined Oracle+Diffusor circuit constructed in the prior steps $t$ times to amplify the probability of measuring the target state. 
    For an $N$-qubit $U_f$ with a unique target, $t = \floor{\pi/4 \cdot \sqrt{N}}$; for $U_f$ with target degeneracy $s$, $t = \floor{\pi/4 \cdot \sqrt{N/s}}$
    
    \item \textbf{Measurement:}
    Measure the entire qubit register. The target state should be observed with high probability.
\end{enumerate}

\subsection{Simulating Circuit Execution: Noise Models}
\label{subsec:noise_modeling_simulations}

As discussed in Section~\ref{sec:background}, simulations of realistic quantum computations must incorporate the effects of noise and decoherence. In our simulations, we model noise at the pulse level by incorporating collapse operators (Lindblad operators) directly into the time evolution of the quantum state. This enables a dynamic, continuous-time representation of errors to capture underlying physical processes (e.g., amplitude damping, dephasing) and their observable effects (e.g., bit-flips, phase-flips). The noise model employed here extends the theoretical framework discussed previously\footnote{Implementation details are openly available in the \texttt{NoiseModel} class, lines 27-57
define the creation and inclusion of multiple collapse operators for amplitude damping, dephasing, and discrete error mechanisms.} . Each qubit is subject to relaxation and dephasing with characteristic times \(T_1\) (amplitude damping) and \(T_2\) (dephasing). In addition, we incorporate discrete error mechanisms using Kraus operators through effective collapse operators. These include:

\begin{itemize}
    \item \textbf{Bit-flip errors} implemented via \(\sigma_x\), flipping \(|0\rangle \leftrightarrow |1\rangle\) with a given probability.
    \item \textbf{Phase-flip errors} implemented using \(\sigma_z\), introducing relative phases that diminish coherence.
    \item \textbf{Bit-phase-flip and depolarizing noise:} obtained by combining Pauli operators \(\sigma_x\), \(\sigma_y\), and \(\sigma_z\), modeling isotropic and more complex error processes.
\end{itemize}

By embedding these collapse operators into the Lindblad master equation, each simulation run accounts for energy relaxation, dephasing, and probabilistic gate errors at every time step of the pulse-level evolution. This continuous and integrated noise modeling is more realistic and fine-grained than simple gate-level noise approximations. It captures the interplay between control pulses, system Hamiltonian, and environment-induced decoherence. As the genetic optimization algorithm iteratively modifies pulse parameters, it naturally adapts to mitigate these dynamical noise processes. This strategy aligns with the NISQ philosophy: rather than relying solely on idealized gate-level abstractions, we address noise directly at the physical control layer, refining pulses to improve overall fidelity.

\subsection{Simulation Parameters}

To conduct our numerical experiments and evolutionary optimization, we established a standard set of simulation parameters. We aimed to demonstrate proof of principle with representative instances while remaining within reasonable computational requirements for broad external replication, given the CPU intensity of the process.

\begin{itemize}
    \item \textbf{Number of Qubits:}  
    We focus on four-qubit (Deutsch-Jozsa) and four-qubit (Grover) systems. This choice balances complexity and feasibility, enabling non-trivial benchmarks while remaining computationally tractable.

    \item \textbf{Initial and Target States:}  
    Each benchmark algorithm defines an initial state \(|000\ldots0\rangle\) and a known target state. For Deutsch-Jozsa, the final measurement pattern distinguishes constant from balanced functions. For Grover, the target state \(|1111\rangle\) should be measureable after applying the oracle and diffusion steps. The fidelity is calculated between the final simulated state and this ideal target state.

    \item \textbf{Noise Parameters:}  
    Noise is parameterized by \(T_1\), \(T_2\), and error probabilities (bit-flip, phase-flip). These values emulate typical conditions of NISQ devices \cite{Greenbaum_2017,ma2024reshapingquantumdevicenoise, PhysRevLett.131.200602, PhysRevA.94.052325} .  In this study, we concentrate on a \textbf{combined noise regime} that integrates amplitude damping, dephasing, and independent Pauli errors. This focus is justified by the current stage of our research, which aims to develop and validate the foundational aspects of our adaptive genetic algorithm. Exploring multiple and more varied noise regimes introduces additional complexities that are reserved for subsequent investigations. By adjusting them, we can evaluate the robustness of the optimization algorithm under various noise regimes.

    \item \textbf{Pulse and Gate Parameterization:}  
    Quantum gates are synthesized into parameterized pulses defined by their durations, amplitudes, and phases. Our genetic algorithm specifically optimizes the evolution time (\texttt{evo\_time}) and the number of time slices (\texttt{num\_tslots}) for each pulse, as outlined in Section~\ref{sec:qutip-qip}. Simultaneously, each simulation performs a comprehensive time-dependent evolution under the specified Hamiltonian, control field and noise channels, ensuring that the refined pulse parameters enhance gate fidelity within noisy environments.

    \item \textbf{Genetic Algorithm Configuration:}  
    For both the Deutsch-Jozsa and Grover's algorithms, the genetic algorithm was configured to run for 250 generations with a population size of 500 individuals. The evaluation of fitness functions was distributed across available CPU cores. This parallelization was achieved using the  Scalable Concurrent Operations in Python (SCOOP), library ~\cite{scoop}, which facilitates dynamic load balancing and ensures efficient utilization of computational resources. This strategy significantly reduced time and enhanced the scalability of the simulations, enabling more extensive and accurate experimentation. While SCOOP effectively streamlines parallelization, potential overheads may arise from task scheduling and inter-process communication. For example, the broker-worker model may introduce slight delays in dispatching tasks, especially when the number of tasks exceeds the number of available resources. Similarly, serialization and data transfer between processes can contribute to overhead, particularly when working with large datasets or deeply nested task hierarchies. Measuring overhead in the process was out of scope for the work presented here.

    \item \textbf{Solver Options and Convergence:}  
    We use QuTiP’s \texttt{mesolve} with a high \(\texttt{nsteps}\) limit (e.g., 100{,}000) to ensure accurate time evolution under complex pulse profiles. This parameter sets the maximum number of internal integration steps, letting QuTiP’s solver \cite{lambert2024qutip5quantumtoolbox} refine the time grid as needed when pulse variations or collapse operators change rapidly. Although the solver may not use all 100{,}000 steps, having this high upper bound avoids integration failures or skipping 
    important pulse features. Convergence was tracked through stable fidelity values and an early-stopping criterion in the genetic algorithm

    \item \textbf{Data Logging and Visualization:}  
    Throughout the optimization process, we log fidelity values, best solutions, and parameter evolutions to CSV files. Visualizations include pulse shapes, fidelity evolution curves, and histograms of parameters, enabling comprehensive post-analysis.
\end{itemize}

\begin{table*}[ht!]
\centering
\scriptsize
\caption{Parameters for circuit simulations with noise (Short and Long Runs).}
\label{tab:doe_and_longtrun}
\begin{minipage}{0.48\textwidth}
\centering
\caption*{(a) Short Runs}
\begin{tabular}{l r}
\toprule
\textbf{Parameter} & \textbf{Value} \\
\midrule
Circuits & Deutsch-Jozsa, Grover \\ 
GA Population & 50 \\
Generations & 30 \\
$T_1$ & 50.0 \\
$T_2$ & 30.0 \\
Prob. bit flip & 0.02 \\
Prob. phase flip & 0.02\\
\bottomrule
\end{tabular}
\end{minipage}%
\hfill
\begin{minipage}{0.48\textwidth}
\centering
\caption*{(b) Long Runs}
\begin{tabular}{l r}
\toprule
\textbf{Parameter} & \textbf{Value} \\
\midrule
Circuits & Deutsch-Jozsa, Grover \\ 
GA Population & 250 \\
Generations & 500 \\
$T_1$ & 50.0 \\
$T_2$ & 30.0 \\
Prob. bit flip & 0.02 \\
Prob. phase flip & 0.02\\
\bottomrule
\end{tabular}
\end{minipage}
\end{table*}

The corresponding design of experiments is captured in Table \ref{tab:doe_and_longtrun}.

\section{Results}
\label{sec:results}

In this section, we present numerical outcomes for two illustrative quantum algorithms:
\textbf{Deutsch-Jozsa} (Section~\ref{subsec:deutschjozsa}) and \textbf{Grover}
(Section~\ref{subsec:grover_results}). Our experiments rely on a genetic pulse-level search configured with a population size of 500 and run for 250 generations, as detailed in the long-run setup (Table~\ref{tab:doe_and_longtrun} (b)) testing with 2,3, and 4 qubits per algorithm. To evaluate the variability across various parameter configurations, we also conducted short-run experiments with a population size of 50, 15 generations, and 4 qubits per algorithm (Table~\ref{tab:doe_and_longtrun} (a)). These short runs serve as a comparative baseline to characterize the impact of altering the genetic algorithm parameters on the performance of the quantum algorithms.

\subsection{Deutsch-Jozsa Algorithm}
\label{subsec:deutschjozsa}

We begin with the Deutsch-Jozsa algorithm, evaluated over 250 generations with a population size of 500 within a genetic pulse-level procedure (Table~\ref{tab:doe_and_longtrun} (a)) and using 4 qubits\footnote{Detailed results: \texttt{DeutschJozsa\_4Q\_With\_Opt\_log.csv}. See \textit{Code and Data Availability} below for data access and replication.} . In addition, we executed the algorithm with various qubit configurations. The results are presented in Table~\ref{tab:deutschjozsa_experiments}. We recorded the \texttt{qubits}, \texttt{original fidelity average}, \texttt{optimized fidelity maximum}, \texttt{optimized fidelity generation}, and \texttt{total execution time}.

\subsubsection{Fidelity Evolution and Pulse Representation}
\label{subsubsec:dj_fidelity_evolution}

\begin{figure}[h!]
\centering
\scalebox{0.9}{
\begin{tikzpicture}
\begin{axis}[
    width=\linewidth,
    xlabel={Generation},
    ylabel={Fidelity $\pm$ std},
    grid=both,
    title={Deutsch-Jozsa Algorithm: Fidelity Over Generations},
    legend style={at={(0.03,0.97)},anchor=north west},
]
\addplot+[
    mark=o,
    mark size=1.5pt,
    thick,
    color=blue,
    error bars/.cd,
        y dir=both,
        y explicit,
] table [
    x=gen,
    y=avg,
    y error=std,
    col sep=comma
] {DeutschJozsa_4Q_With_Opt_log.csv};
\addlegendentry{avg $\pm$ std}
\end{axis}
\end{tikzpicture}
}
\caption{Average fidelity (\texttt{avg}) and standard deviation (\texttt{std}) for the Deutsch-Jozsa algorithm over 250 generations of pulse-level genetic optimization. The optimization process required approximately 1 hour and 30 minutes to complete using 6 cores. See Table \ref{tab:vm_specs} for CPU specifications. See Table \ref{tab:deutschjozsa_experiments} experiment with 4 qubits for more details with an early stop around before generation number 80.} 
\label{fig:dj_avg_std}
\end{figure}
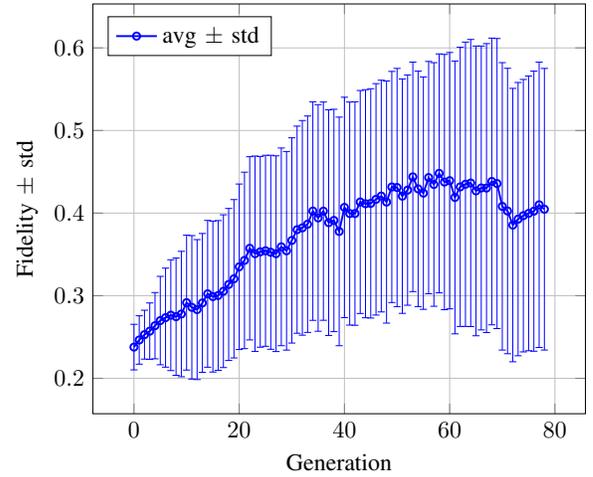

Figure~\ref{fig:dj_avg_std} evidences the change in average fidelity and standard deviation given the pulse choices made by the evolutionary algorithm with each generation. The maximum average fidelity of 0.4481 was achieved at generation 58. The overall highest fidelity (i.e., the best single individual in any generation) was 0.6167, first reached at generation 55. The simulation had an early stop on generation 78 due to the absence of a substantial improvement over the preceding 20 generations.

The standard deviation (\texttt{std}) in fidelity grows from approximately 0.028 at the outset (generation 0) to about 0.170 by generation 78. Early in the run, a relatively low standard deviation indicates that many individuals share similar (though modest) fidelity levels. As evolution proceeds and the genetic algorithm explores a broader swath of the parameter space, the standard deviation increases. Ultimately, this moderate spread (~0.17) at the final generation implies that, although the population has converged on higher-fidelity solutions, it still maintains some variability in pulse parameters.

\begin{figure*}[h!]
\centering
\includegraphics[width=\linewidth]{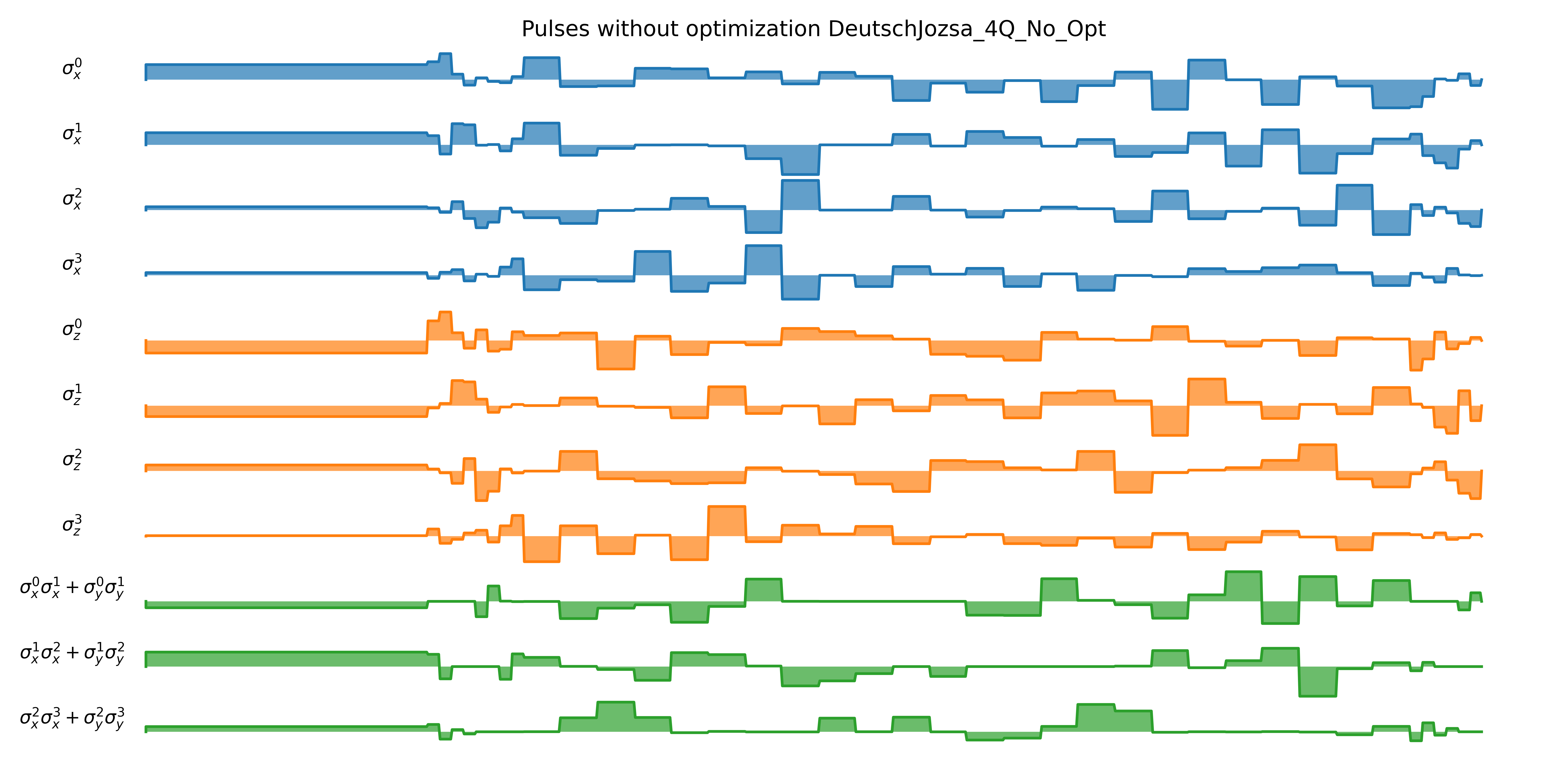}
\caption{Baseline pulse waveforms for Deutsch-Jozsa obtained for the best-performing individual across all simulations.}
\label{fig:dj_pulses_base}
\end{figure*}

\begin{figure*}[h!]
\centering
\includegraphics[width=\linewidth]{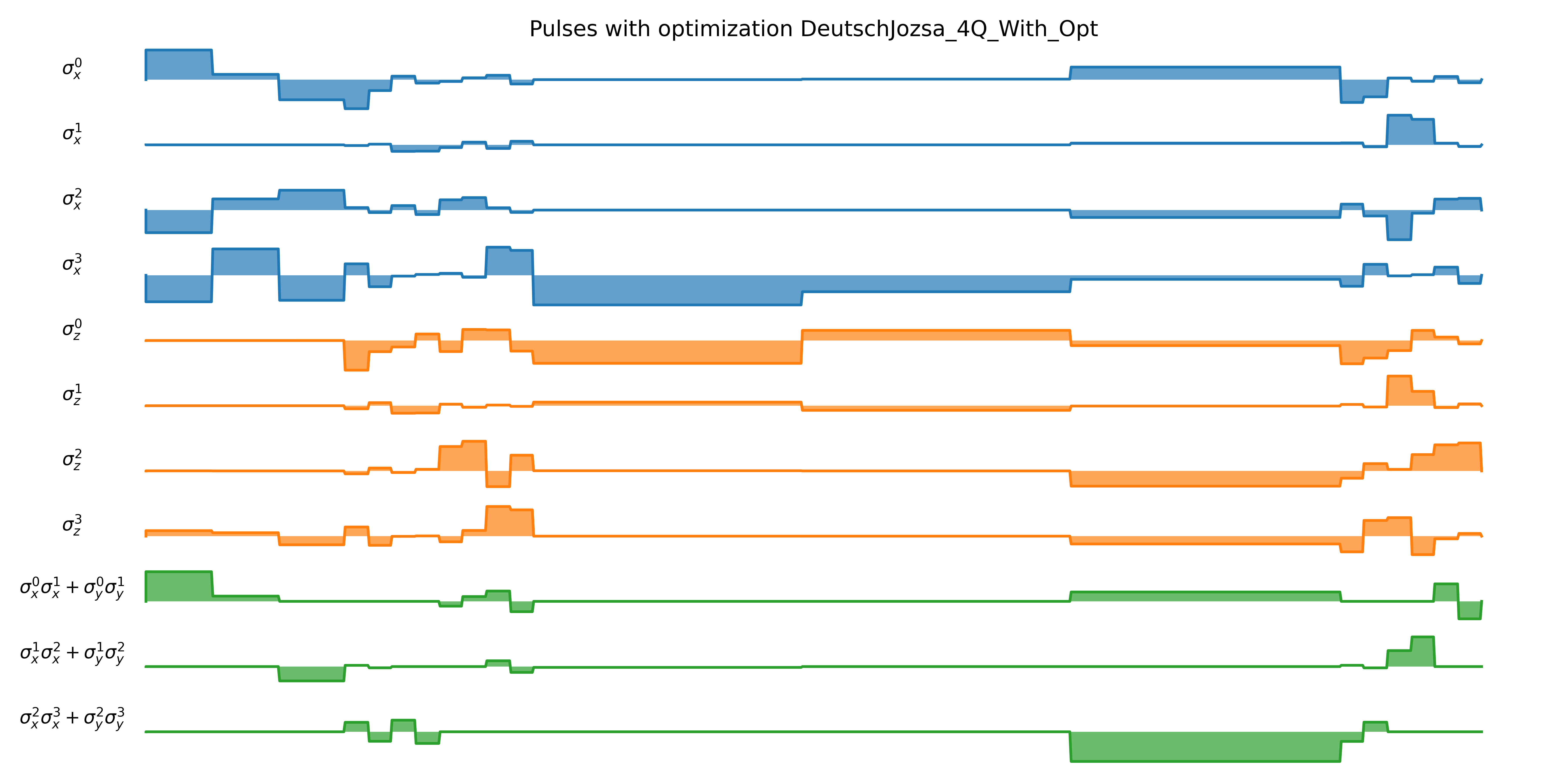}
\caption{Optimized pulse waveforms for Deutsch-Jozsa obtained for the best-performing individual across all simulations..}
\label{fig:dj_pulses}
\end{figure*}

Figure~\ref{fig:dj_pulses} captures the final optimized waveforms from the best performing run. We observe amplitude profiles that differ substantially from those generated using QuTiP default functions.

\subsubsection{Results: Baseline vs.\ Optimized}
\label{subsubsec:dj_fidelity_comparison}

\begin{figure}[h!]
\centering
\begin{tikzpicture}
\begin{axis}[
    ybar,
    bar width=0.3cm,
    width=0.7\linewidth,
    height=0.5\linewidth,
    enlarge x limits=0.5,
    symbolic x coords={DeutschJozsa\_4Q},
    xtick=data,
    ylabel={Fidelity},
    ymin=0,
    ymax=0.9,
    grid=both,
    title={Fidelity Comparison: Deutsch--Jozsa (4Q)},
    legend style={
        at={(0.5,1.05)},
        anchor=south,
        legend columns=2
    },
    nodes near coords,
    every node near coord/.append style={font=\tiny},
]
\addplot coordinates { (DeutschJozsa\_4Q, 0.2367) };
\addplot coordinates { (DeutschJozsa\_4Q, 0.4480) };
\addplot coordinates { (DeutschJozsa\_4Q, 0.6167) };
\legend{No Optimization, Optimized, Max Fidelity}
\end{axis}
\end{tikzpicture}
\caption{Fidelity outcomes for the Deutsch-Jozsa algorithm, baseline vs. optimized using our genetic algorithm. Values correspond to the average fidelity obtained after 58 generations (With Optimization) and the maximum fidelity obtained after 58 generations. An improvement of 0.2113 is observed in average fidelity and 0.38 in maximum fidelity, simulation had an early stop on generation 78 due to the absence of a substantial improvement over the preceding 20 generations.}
\label{fig:dj_bar_comparison}
\end{figure}
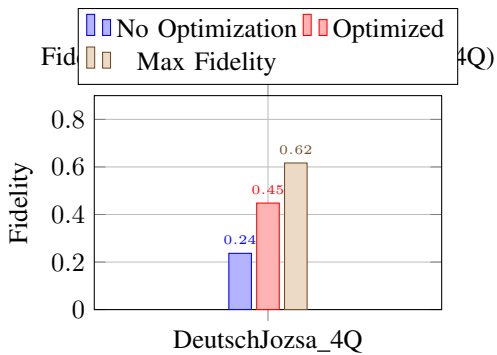

Figure~\ref{fig:dj_bar_comparison} illustrates a substantial net enhancement (\(+0.2113\))  on average when pulses evolve under the genetic method. The long-run data (\texttt{gen}=0 to 249) reveal how average fidelity can climb from approximately 0.23 to beyond 0.61. 

\begin{table*}[t!]
\centering
\scriptsize
\caption{Final Outcomes of Deutsch-Jozsa Experiments 4 qubits (Short Runs)}
\label{tab:deutschjozsa_experiments}
\begin{tabular}{l r r r r}
\toprule
\textbf{Exp.} & \textbf{Baseline Fidelity} & \textbf{Avg. Optimized Fidelity} & \textbf{Max. Optimized Fidelity} & \textbf{Generation}\\
\midrule
1 & 0.2883 & 0.2797 & 0.3150 & 10 \\
2 & 0.2614 & 0.2720 & 0.2842 & 10 \\
3 & 0.2536 & 0.2652 & 0.2812 & 12\\
4 & 0.1654 & 0.2995 & 0.3679 & 14\\
5 & 0.2729 & 0.2755 & 0.2992 & 7\\
\midrule
\textit{Mean} & 0.2483 & 0.2784 & 0.3095 & \\
\textit{SD} & 0.0453 & 0.0114 & 0.0321 & \\
\bottomrule
\end{tabular}
\end{table*}

\paragraph{Short Experiments (15 Generations)}

Table~\ref{tab:deutschjozsa_experiments} presents the results of five additional short-run experiments on the 4‑qubit Deutsch-Jozsa algorithm, each with a population size of 50 and limited to 15 generations. Among these, four experiments (Experiments~2,~3,~4, and~5) demonstrated improvements in average optimized fidelity, with average fidelities increasing from a mean baseline of 0.2483 to 0.2784. Notably, Experiment~4 exhibited the most significant enhancement, boosting its average fidelity from 0.1654 to 0.2995 and achieving a maximum optimized fidelity of 0.3679. Conversely, Experiment~1 experienced a slight decrease in average fidelity from 0.2883 to 0.2797, product of the inherent variability and stochastic nature of genetic algorithms.

\subsection{Grover’s Algorithm}
\label{subsec:grover_results}

We evaluated Grover’s algorithm over 250 generations with a population size of 500 within a genetic pulse-level procedure (Table~\ref{tab:doe_and_longtrun} (b)) and using 4 qubits \footnote{Detailed results: \texttt{Grover\_4Q\_With\_Opt\_log.csv}. See \textit{Code and Data Availability} below for data access and replication.} In addition, we executed the algorithm with various qubit configurations. The results are presented in Table~\ref{tab:grover_experiment_summary}. We recorded the \texttt{qubits}, \texttt{original fidelity average}, \texttt{optimized fidelity maximum}, \texttt{optimized fidelity generation}, and \texttt{total execution time}.

\subsubsection{Fidelity Evolution and Pulse Representation}

\label{subsubsec:grover_fidelity_evolution}

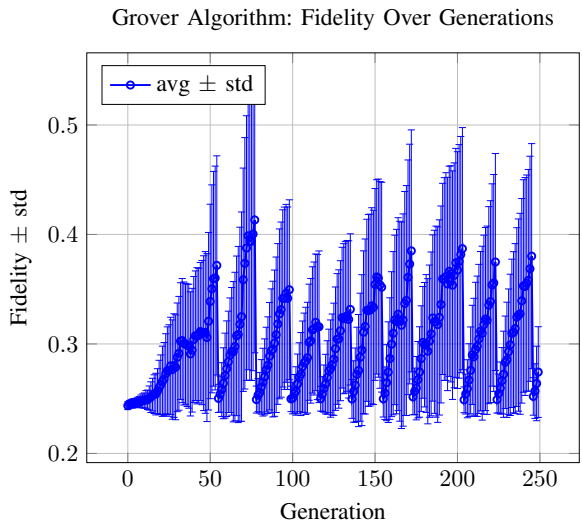
\begin{figure}[h!]
\centering
\scalebox{0.9}{
\begin{tikzpicture}
\begin{axis}[
    width=\linewidth,
    xlabel={Generation},
    ylabel={Fidelity $\pm$ std},
    grid=both,
    title={Grover Algorithm: Fidelity Over Generations},
    legend style={at={(0.03,0.97)},anchor=north west},
]
\addplot+[
    mark=o,
    mark size=1.5pt,
    thick,
    color=blue,
    error bars/.cd,
        y dir=both,
        y explicit,
] table [
    x=gen,
    y=avg,
    y error=std,
    col sep=comma
] {Grover_4Q_With_Opt_log.csv};
\addlegendentry{avg $\pm$ std}
\end{axis}
\end{tikzpicture}
}
\caption{Average fidelity and standard deviation for Grover’s algorithm,
plotted across 250 generations of genetic optimization. The entire optimization process required approximately 16 hours to complete using 8 coresas per Table \ref{tab:grover_experiment_summary}. No early stop was implemented in this case, which led to a substantial increase in compute time.}
\label{fig:grover_avg_std}
\end{figure}

\begin{figure*}[h!]
\centering
\includegraphics[width=\linewidth]{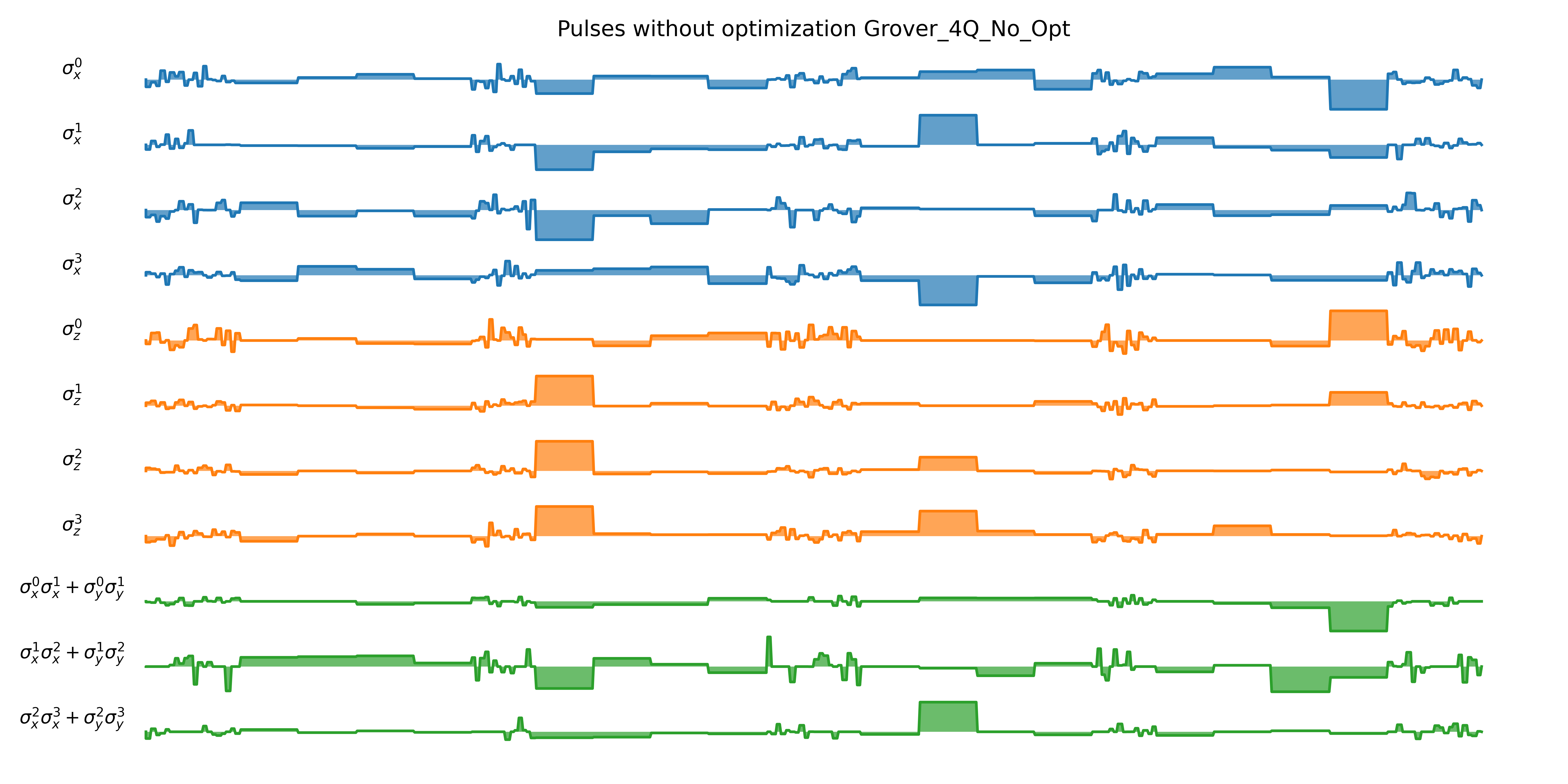}
\caption{Baseline pulse waveforms for Grover’s algorithm obtained for representative individual across all simulations.}
\label{fig:grover_pulses_base}
\end{figure*}

\begin{figure*}[h!]
\centering
\includegraphics[width=\linewidth]{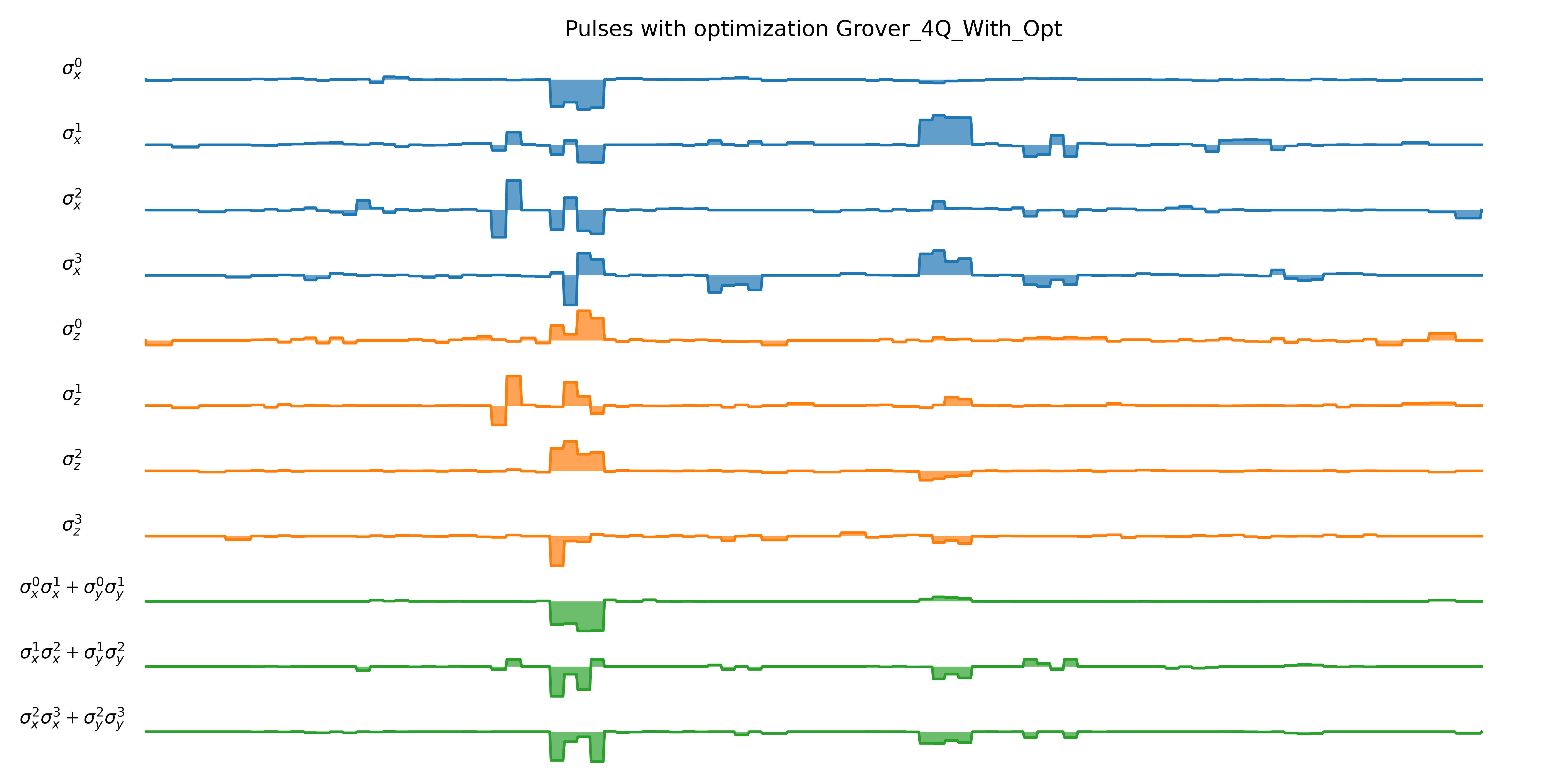}
\caption{Optimized pulse waveforms for Grover’s algorithm obtained for the same representative individual across all simulations.}
\label{fig:grover_pulses}
\end{figure*}

Figures~\ref{fig:grover_avg_std} and \ref{fig:grover_pulses} capture
changes in fidelity with pulses chosen by the genetic algorithm for Grover’s algorithm, and pulse waveforms for the best performing individual. The genetic search occasionally induces short-term drops when new candidates are introduced but can ultimately identify pulses that surpass naive gate-level baselines.

\subsubsection{Results: Baseline vs. Optimized}
\label{subsubsec:grover_fidelity_comparison}

\begin{figure}[h!]
\centering
\begin{tikzpicture}
\begin{axis}[
    ybar,
    bar width=0.3cm,
    width=\linewidth,
    height=0.5\linewidth,
    enlarge x limits=0.5,
    symbolic x coords={Grover\_4Q},
    xtick=data,
    ylabel={Fidelity},
    ymin=0,
    ymax=0.6,
    grid=both,
    title={Fidelity Comparison: Grover (4Q)},
    legend style={
        at={(0.5,1.05)},
        anchor=south,
        legend columns=3
    },
    nodes near coords,
    every node near coord/.append style={font=\tiny},
]
\addplot coordinates { (Grover\_4Q, 0.2445) };
\addplot coordinates { (Grover\_4Q, 0.3870) };
\addplot coordinates { (Grover\_4Q, 0.4907) };
\legend{No Optimization, Final Optimized, Max Fidelity}
\end{axis}
\end{tikzpicture}
\caption{Fidelity outcomes for the Grover algorithm, baseline vs. optimized using our genetic algorithm. Values correspond to the average fidelity obtained after 203 generations (Final Optimized) and the maximum fidelity obtained after 203 generations. An improvement of 0.1425 is observed in average fidelity and 0.2462 in maximum fidelity.}
\label{fig:grover_bar_fidelity_comparison}
\end{figure}
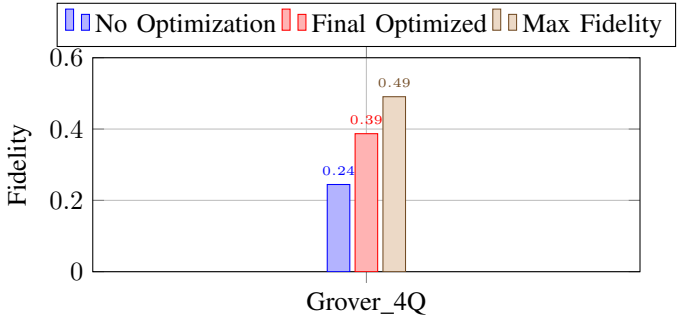

Figure~\ref{fig:grover_bar_fidelity_comparison} highlights a representative run where the optimized pulses achieved significant improvements over the baseline. The results emphasize the effectiveness of the genetic algorithm, with the final optimized fidelity showing an average increase of 0.1425 and the maximum fidelity improving by 0.2462 after 203 generations. These outcomes also underscore the stochastic nature of the algorithm, as improvements depend on the evolutionary process and can vary between runs.Table~\ref{tab:grover_experiment_summary}
reports outcomes for multiple short-run experiments:

\begin{table*}[t!]
\centering
\scriptsize
\caption{Final Outcomes of Grover Experiments 4 qubits (Short runs)}
\label{tab:grover_experiment_summaryshort}
\begin{tabular}{l r r r r}
\toprule
\textbf{Exp.} & \textbf{Original Fidelity} & \textbf{Avg. Optimized Fidelity} & \textbf{Max. Optimized Fidelity} & \textbf{Generation}\\
\midrule
1 & 0.2436 & 0.2440 & 0.2511 & 3\\
2 & 0.2500 & 0.2476 & 0.2535 & 1\\
3 & 0.2461 & 0.2510 & 0.3311 & 4\\
4 & 0.2474 & 0.2496 & 0.2574 & 14\\
5 & 0.2403 & 0.2691 & 0.3838 & 9\\
\midrule
\textit{Mean} & 0.2455 & 0.2523 & 0.2954 &\\
\textit{SD} & 0.0035 & 0.0101 & 0.0545 & \\
\bottomrule
\end{tabular}
\end{table*}

\begin{table*}[t!]
\centering
\scriptsize
\caption{Final Outcomes of Deutsch-Jozsa Experiments (Long Runs)}
\label{tab:dj_experiment_summary}
\begin{tabular}{l r r r r r r}
\toprule
\textbf{Algorithm} & \textbf{Qubits} & \textbf{Original Fidelity} & \textbf{Avg. Optimized Fidelity} & \textbf{Max. Optimized Fidelity} & \textbf{Generation} & \textbf{Total Execution Time} \\
\midrule
Deutsch-Jozsa & 2 & 0.4251 & 0.5376 & 0.6843 & 17 & 159.40 seconds \\
Deutsch-Jozsa & 3 & 0.3457 & 0.5509 & 0.6608 & 56 & 1610.31 seconds \\
Deutsch-Jozsa & 4 & 0.2367 & 0.4480 & 0.6167 & 58 & 4981.08 seconds \\
\midrule
\textit{Mean} &  & 0.3358 & 0.5122 & 0.6539 & -- & -- \\
\textit{SD} &  & 0.0946 & 0.0560 & 0.0343 & -- & -- \\
\bottomrule
\end{tabular}
\end{table*}

\begin{table*}[t!]
\centering
\scriptsize
\caption{Final Outcomes of Grover Experiments (Long Runs)}
\label{tab:grover_experiment_summary}
\begin{tabular}{l r r r r r r}
\toprule
\textbf{Algorithm} & \textbf{Qubits} & \textbf{Original Fidelity} & \textbf{Avg. Optimized Fidelity} & \textbf{Max. Optimized Fidelity} & \textbf{Generation} & \textbf{Total Execution Time} \\
\midrule
Grover & 2 & 0.4943 & 0.7218 & 0.8473 & 79 & 1224.59 seconds \\
Grover & 3 & 0.3417 & 0.6184 & 0.7648 & 76 & 5268.40 seconds \\
Grover & 4 & 0.2444 & 0.3870 & 0.4907 & 203 & 57600.40 seconds \\
\midrule
\textit{Mean} &  & 0.3601 & 0.5757 & 0.7009 & -- & -- \\
\textit{SD} &  & 0.1029 & 0.1400 & 0.1524 & -- & -- \\
\bottomrule
\end{tabular}
\end{table*}

Experiments~1,~3,~4, and~5 exceed the baseline fidelities, whereas Experiment~2 does not. These outcomes indicate that, for most runs, the genetic algorithm successfully navigated the parameter space to identify more optimal pulse configurations leading to enhanced fidelities. Specifically, Experiments~1,~3,~4, and~5 demonstrated improvements in both average and maximum fidelities, evidence of the ability of the algorithm to discover effective solutions. Conversely, Experiment~2 suggests that the algorithm may have converged prematurely to a local maximum or experienced insufficient exploration, hindering significant fidelity enhancements. This variability underscores the importance of robust exploration strategies and adequate parameter settings to ensure consistent performance across different runs. 

\section{Discussion}
\label{sec:discussion}

We now turn our attention to the analysis of findings obtained from the experiments described above. For each algorithm, we concentrate on a) the outcomes of long experiments, b) outcomes from short experiments, and c) drawing a consolidated veredict of performance. We then provide a synthesis of our current understanding and a description of future directions based on these results, with an emphasis on the potential and limitations present in our algorithm.

\subsection{Deutsch-Jozsa}
\label{subsec:dj_observations}

\paragraph{Long Experiments (250 generations).}
Table \ref{tab:dj_experiment_summary} summarizes the final outcomes of the 2‑, 3‑, and 4‑qubit Deutsch-Jozsa experiments conducted over long runs (up to 250 generations). Before optimization, the original fidelity decreases from 0.4251 (2 qubits) down to 0.2367 (4 qubits), reflecting the added complexity in implementing a larger quantum circuit at the pulse level. After the genetic algorithm is applied, we observe sizable enhancements in the average optimized fidelity for all system sizes—most notably increasing from 0.2367 to 0.4480 for the 4‑qubit setting. A closer look at the maximum optimized fidelity column reveals that the best individual solutions achieve even higher fidelities, with values ranging from 0.6167 (4 qubits) up to 0.6843 (2 qubits). These maximum fidelities demonstrate the algorithm’s capability to discover high‑performing pulse configurations, even though the overall average fidelity may be somewhat lower.

Our results show the number of generations required for these improvements varies across different qubit counts. For the 2‑qubit system, a relatively modest 17 generations sufficed to reach a final plateau, taking roughly 159 seconds total. By contrast, the 4‑qubit system required 58 generations and approximately 5,000 seconds, underscoring how larger quantum systems demand substantially more computation time—both for pulse-level simulation and for searching a more complex parameter landscape. These findings confirm that our genetic algorithm reliably boosts fidelity compared to the original (unoptimized) pulses even for larger circuits. As shown in the bottom rows, the mean original fidelity (across all system sizes) is 0.3358, which rises to 0.5122 after optimization and 0.6539 at its best single-individual performance. The improvement is both consistent and statistically significant, reflecting the robustness of evolutionary techniques for pulse-level control in quantum computing.

\paragraph{Short Experiments (30 generations.}
Section~\ref{subsubsec:dj_fidelity_comparison} and
Table \ref{tab:deutschjozsa_experiments} summarizes five short-run experiments (capped at 30 generations) on the 4‑qubit Deutsch-Jozsa algorithm, each terminating early, often before reaching 15 generations—once incremental gains diminished. Despite the truncated optimization time, each trial demonstrates  at least modest improvements over the baseline fidelity. Across the five experiments, the baseline fidelity varies from 0.1654 in Experiment 4 to 0.2883 in Experiment 1, with a mean of approximately 0.2483. After allowing the genetic algorithm to evolve pulses for a handful of generations, the average optimized fidelity rises in four out of five experiments, reaching values between 0.2652 (Exp 3) and 0.2995 (Exp 4). On average, the improvement lifts fidelity from 0.2483 to 0.2784—a net gain of about 0.03 in average fidelity. For individual best solutions, the maximum optimized fidelity follows a similarly encouraging pattern. Experiment 4 shows the largest jump: from a baseline of only 0.1654 to a maximum fidelity of 0.3679. In most other experiments, the max fidelity plateaus closer to 0.28–0.32, still exceeding the original baseline by a meaningful margin.

Taken together, these results suggest that even within a short time frame (7 to 14 generations), genetic optimization can discover nontrivial pulse sequences that exceed the baseline performance. The relatively modest increases in certain experiments (e.g., Exp 1, Exp 2, and Exp 5) can be attributed to both the complexity of the 4‑qubit circuit and the limited evolutionary horizon (fewer generations). Nonetheless, the consistency of gains across all trials underscores the potential of our algorithm to offer quick improvements without the computational investment of a longer, 250‑generation run.

\paragraph{Consolidated Verdict}
Our experiments demonstrate the capacity of pulse-level genetic optimization to systematically improve fidelity in the Deutsch-Jozsa algorithm with noise across a range of qubit counts and run lengths. In the \emph{long} (250‑generation) experiments, average fidelity increases substantially over baseline—even in the more demanding 4‑qubit scenario—reaching up to 0.4480 on average and 0.6167 in the best individual solution. These gains come at the cost of a heavier computational load, as evidenced by the 2‑qubit case converging within 17 generations while the 4‑qubit instance required over 50 generations and significantly longer wall-clock time.  By contrast, the \emph{short} (30‑generation) experiments underline how measurable improvements can still be secured on a tighter schedule: despite capping runs at under 15 generations in most trials, the 4‑qubit baseline fidelity of around 0.25 was consistently lifted to the high 0.27–0.30 range on average, and up to 0.37 for the best individual. Although these short runs deliver more modest enhancements compared to the long-run studies, they highlight a notable “quick-gain” aspect of genetic optimization in pulse-level quantum control.

These findings confirm that evolutionary algorithms are robust for incrementally refining quantum gate implementations at the pulse level. Whether time and resources permit a thorough exploration of hundreds of generations or merely a handful, the genetic algorithm consistently uncovers pulse sequences that outperform fidelities in the pulse-level baseline.

\subsection{Grover Observations}
\label{subsec:grover_observations}

\paragraph{Long Experiments (250 generations).} Table \ref{tab:grover_experiment_summary} summarizes the final outcomes of the 2‑, 3‑, and 4‑qubit Grover experiments conducted over long runs (up to 250 generations). Prior to optimization, the original fidelity decreases from 0.4943 (2 qubits) down to 0.2444 (4 qubits), showing the effects of increasing complexity associated with scaling the quantum circuit at the pulse level. Upon applying the genetic algorithm, significant improvements are observed in the average optimized fidelity across all qubit configurations. Specifically, for the 2‑qubit system, the average fidelity increases from 0.4943 to 0.7218 while the 4‑qubit system sees an increase from 0.2444 to 0.3870. Examining the maximum optimized fidelity reveals even more substantial gains. The 2‑qubit system reaches a maximum fidelity of 0.8473, up from the original 0.4943, and the 4‑qubit system achieves a maximum fidelity of 0.4907, compared to its baseline of 0.2444. Once again, the algorithm shows promising results.

The number of generations required to attain these improvements varies as a function of the number of qubits. The 2‑qubit experiment converges after 79 generations, taking a total execution time of approximately 1,224.59 seconds. In contrast, the 4‑qubit experiment requires 203 generations and significantly more computational time, totaling 57,600.40 seconds. This is consistent with the increased computational demands and complexity involved in optimizing larger quantum systems. At the same time, our genetic algorithm consistently enhances fidelity across different qubit counts. The mean original fidelity across all tested systems is 0.3601, which rises to an average optimized fidelity of 0.5757 and reaches a mean maximum fidelity of 0.7009. The standard deviations (SD) of 0.1029, 0.1400, and 0.1524 for the original, average optimized, and maximum optimized fidelities, respectively, indicate a reliable and statistically significant improvement. These results affirm the robustness of evolutionary strategies in optimizing pulse-level control for quantum algorithms, even as circuit complexity increases.

\paragraph{Five Short-Run Experiments.}
From Table~\ref{tab:grover_experiment_summaryshort} presents the results of five short-run experiments on the 4‑qubit Grover algorithm, each utilizing a population size of 50 and limited to 15 generations. Among these, four experiments (Experiments~1,~3,~4, and~5) demonstrated improvements in average optimized fidelity, with average fidelities increasing from a mean baseline of 0.2455 to 0.2523. Experiment~5 exhibited the most significant enhancement, boosting its average fidelity from 0.2403 to 0.2691 and achieving a maximum optimized fidelity of 0.3838 after 9 generations. Conversely, Experiment~2 experienced a slight decrease in average fidelity from 0.2500 to 0.2476. We suspect that certain trajectories leading to local maxima at fidelities lower than the original may occur during the exploratory phase of the algorithm. In consequence, and by virtue of the approximate modeling of continuous-time processes, the latter suggests first-passage time theory \cite{maniscalco2004lindblad} may help better understand the underlying landscape. The standard deviations (SD) provide evidence of the consistency of the improvements, with the average optimized fidelity showing an SD of 0.0101 and the maximum optimized fidelity exhibiting an SD of 0.0545. These findings confirm that even within limited generations, the genetic algorithm can effectively enhance fidelity for the Grover algorithm notwithstanding the occasional decrease in fidelity for reasons explained above.

\paragraph{Consolidated Verdict}
These results confirm that the genetic algorithm robustly improves pulse-level fidelity for Grover’s algorithm across different qubit sizes. In the long-run experiments, even when baseline fidelity declines significantly for higher qubit counts, the genetic algorithm consistently improves both the average and maximum fidelities by non-trivial margins (e.g., from 0.2444 to 0.3870 on average for the 4‑qubit system). Meanwhile, short-run tests (15 generations) reveal that modest yet meaningful fidelity gains are achievable within a limited evolutionary horizon, despite occasional decreases in certain runs. These fluctuations align with the stochastic nature of genetic algorithms, where convergence can be heavily influenced by random initializations and population diversity. Taken together, these findings validate the efficacy of evolutionary techniques in pulse-level quantum control, offering a practical path to enhance fidelity --even under constrained runtimes- while also highlighting the need for repeated trials and careful hyperparameter management to account for inherent variability. This particular point delineates an area of future work.

\subsection{Overall Synthesis and Future Directions}
\label{subsec:overall_synthesis}

Evaluating both algorithms suggests the following key observations:

\begin{itemize}
    \item \textbf{Long-Run Gains vs.\ Short-Run Variability:} 
    Extended runs often achieve meaningful improvements (e.g., fidelity for Deutsch--Jozsa and for Grover). Short-run tests
    exhibit a mixture of positive and negative net changes, reinforcing the need for multiple trials and robust population management. We observe across our experiments that increasing circuit depth leads to more pronounced difficulties in mitigating noise; deeper circuits lead to longer evolution times and more persistent, cumulative effects from noise sources. This observation is consistent with fundamental limits on error-mitigation performance reported by Takagi \textit{et al.}~\cite{Takagi_2022}, where the sampling overhead for achieving a target accuracy can grow exponentially with circuit depth. Consequently, although our adaptive genetic algorithm can effectively optimize pulse parameters for shallower circuits, scaling to larger depths within strictly limited resources remains a significant open challenge.

    \item \textbf{Periodic Resets and Exploration:} 
    Momentary dips in average fidelity strongly correlate with soikes in population diversity. This indicates that our algorithm intentionally sacrifices short-term stability to escape local maxima. Such cyclical exploration can lead to eventual breakthroughs but may not succeed in every run with more complex circuits.

    \item \textbf{Parameter Sensitivities:} 
    Mutation rates, crossover probabilities, and replacement tactics
    significantly influence final outcomes. Fine-tuning these
    hyperparameters, or adopting adaptive mutation schedules,
    can stabilize the search and boost fidelity gains more reliably. More research is required to understand the response of the algorithm relative to parameter changes.

    \item \textbf{Next Steps:} 
    Further research might focus on combining pulse-level genetic
    strategies with hardware-specific noise models, exploring synergy with small-scale error-mitigation and testing
    larger circuits. Repeated experiments remain essential to mitigate and testing with other noise regimes
    random-seed fluctuations.
\end{itemize}

Results from our experiments with the Deutsch--Jozsa and Grover algorithms confirm that genetic pulse-level optimization can secure notable fidelity enhancements over default pulses—particularly in longer runs with carefully managed diversity. Even so, occasional low performance correlates with circuit complexity: local maxima, and limited generation counts can all disrupt final outcomes. Despite these challenges, our results show evolutionary methods as a compelling means of refining quantum pulses on Noisy Intermediate-Scale Quantum (NISQ) devices, offering a practical avenue to surpass naive baselines and push algorithmic fidelity closer to theoretical ideals.

Experiments demonstrate that prolonged genetic searches achieve higher fidelities, highlighting the potential to bridge the gap between theoretical models and hardware-constrained implementations. By strategically introducing diversity at specific intervals, the algorithm effectively circumvents stagnation, enabling the discovery of more robust pulse configurations even when initiating from relatively low baseline fidelities. Conversely, shorter runs occasionally underperform compared to the original implementations, underscoring the influence of factors such as insufficient population sizes and suboptimal mutation rates, which can impede the algorithm's ability to escape local maxima. This variability emphasizes the necessity for multiple trials and meticulous parameter tuning, as certain hyperparameter settings may lead to regressions rather than improvements. Despite these challenges, the genetic search method exhibits significant promise: with adequate computational resources, thoughtful management of hyperparameters, and preparedness for occasional setbacks, consistently surpassing fidelities given by default pulse implementations becomes likely.

In summary, the evidence provided in this article indicates that evolutionary pulse-level refinement is practically viable and useful to mitigate noise in contemporary NISQ hardware, provided that researchers remain acutely aware of its potential to deliver significant fidelity enhancements alongside its vulnerability to sporadic declines under suboptimal conditions. We believe that our methodology can help extend the practical capabilities of NISQ devices, thus providing a concrete pathway toward harmonizing real-world device constraints with the theoretical aspirations of quantum algorithms.

\section{Conclusion} \label{sec:conclusion}

We introduced a new adaptive genetic algorithm tailored for pulse-level quantum error mitigation, applying it to prominent quantum algorithms such as Deutsch--Jozsa and Grover' search. Our methodology systematically optimizes pulse parameters across multiple generations, achieving substantial fidelity enhancements  despite occasional fluctuations attributable to diversity-driven resets. These outcomes highlight the efficacy of evolutionary strategies in augmenting quantum circuit performance within noisy environments, all without requiring modifications to the structure and composition of circuits.

Our findings underscore the significance of assisted pulse engineering in bridging the gap between theoretical circuit abstractions and practical hardware implementations. By dynamically evolving parameters that drive control fields, our technique effectively mitigates environmental disturbances and hardware imperfections more efficiently than static gate-level models. This versatile pulse-level framework complements existing error-mitigation and correction methods, extending the functional lifetime of noisy qubits and broadening the scope of feasible NISQ-era computations.

\textbf{Limitations.}
While promising, our technique is not universally guaranteed to surpass baseline fidelities in all scenarios. Certain executions --particularly those with limited population sizes or insufficient generation counts- may prematurely converge to local optima, resulting in suboptimal fidelity outcomes. Tuning these hyperparameters requires further research. Additionally, the reliance on computationally intensive simulations poses scalability challenges, rendering large-scale experiments resource-demanding in both time and computational capacity. This point is particularly salient in the context of large-scale NISQ systems.

\vspace{0.8em}
\noindent
\textbf{Future Work.}
We identify two primary avenues of interest:
\begin{enumerate}
    \item \textbf{Method refinement and integration.} With increased computational capabilities, it becomes feasible to run the algorithm over larger populations and more generations, providing a more comprehensive benchmark across a variety of noise regimes. Additionally, we hypothesize that certain refinements to the GA engine itself (e.g., adaptive mutation schedules) can streamline convergence. Finally, providing integration into existing libraries (e.g., QuTiP) can simplify creating pulse-level optimization workflows for broader adoption.

    \item \textbf{QPU and algorithms testing.} We plan to test our pulse-level optimization strategy on actual quantum hardware to evaluate its ability to perform adaptive error mitigation under real-time noise. Doing this requires access to quantum hardware at the pulse level, opening avenues for collaboration with testbeds and quantum vendors. Testing against a broader range of quantum algorithms, especially those that push NISQ devices to their operational limits, would further confirm the robustness and practical impact of the methods reported here.
\end{enumerate}

\section*{Acknowledgments}

W.A-C. wishes to thank CENFOTEC University for providing direct academic support and resources essential for the completion of this research. S.N-C. thanks the National Center for Supercomputing Applications and the Illinois Quantum Information Science and Technology Center for continued support, and  CENFOTEC University for the opportunity to establish a fruitful research collaboration. Francini Corrales and Geisel Hernández provided insights across the design and execution of the project.Lastly, we acknowledge the QuTiP and qutip-qip teams for their outstanding tools and assistance, which were instrumental in the successful implementation of the quantum simulations reported in this work.

\section*{Code and Data Availability}

All code and scripts required to replicate experiments and results described above are available on our \href{https://github.com/Universidad-Cenfotec/Pulse-Quantum-Error-Correction.git}{GitHub repository}. We provide detailed instructions to facilitate the reproduction of our simulations and analyses on various computational platforms, including environment setup and example scripts.

\bibliographystyle{IEEEtran}
\bibliography{references}

\newpage

\appendix
\label{appendix:algorithm}


\begin{algorithm}
\caption{Population Initialization}
\label{alg:initialization}
\KwData{Population size $N$}
\KwResult{Initial population $P_0$ with $N$ random individuals}

\For{$i \leftarrow 1$ \KwTo $N$}{
    Create individual $ind_i$ with random parameters\;
    Add $ind_i$ to $P_0$\;
}
\Return{$P_0$}\;
\end{algorithm}


\begin{algorithm}
\caption{Parent Selection via Tournament Selection}
\label{alg:selection}
\KwData{Current population $P_g$, Tournament size $k$}
\KwResult{Selected parents for crossover}

Select $k$ individuals randomly from $P_g$\;
Choose the individual with the highest fitness from the selected group as a parent\;
\Return{Selected parents}\;
\end{algorithm}


\begin{algorithm}
\caption{Crossover Operation}
\label{alg:crossover}
\KwData{Pair of parents $(parent_1, parent_2)$, Crossover probability $p_{\text{cross}}$}
\KwResult{Offspring generated through crossover}

\If{random() $<$ $p_{\text{cross}}$}{
    Perform crossover between $parent_1$ and $parent_2$\;
    Create offspring $child_1$ and $child_2$\;
    \Return{$child_1$, $child_2$}\;
}
\Else{
    \Return{$parent_1$, $parent_2$}\;
}
\end{algorithm}


\begin{algorithm}
\caption{Mutation Operation}
\label{alg:mutation}
\KwData{Individual $ind$, Mutation probability $p_{\text{mut}}$}
\KwResult{Mutated individual}

\If{random() $<$ $p_{\text{mut}}$}{
    Mutate the parameters of $ind$\;
}
\Return{$ind$}\;
\end{algorithm}


\begin{algorithm}
\caption{Fitness Evaluation}
\label{alg:fitness_evaluation}
\KwData{Individual $ind$, Quantum circuit model, Noise model}
\KwResult{Fitness value $F(ind)$}

Simulate the quantum circuit using the pulse parameters defined by $ind$\;
Calculate fidelity between the final state and the target state\;
\Return{$F(ind)$}\;
\end{algorithm}


\begin{algorithm}
\caption{Population Replacement}
\label{alg:replacement}
\KwData{Current population $P_g$, Offspring population}
\KwResult{New population $P_{g+1}$}

Replace the worst-performing individuals in $P_g$ with the new offspring\;
\Return{$P_{g+1}$}\;
\end{algorithm}


\begin{algorithm}
\caption{Elitism Strategy}
\label{alg:elitism}
\KwData{Current population $P_g$, Best individual $best\_ind$}
\KwResult{Preserved best individual in $P_{g+1}$}

Add $best\_ind$ to $P_{g+1}$ without modification\;
\Return{$P_{g+1}$}\;
\end{algorithm}


\begin{algorithm}
\caption{Diversity Control Mechanism}
\label{alg:diversity_control}
\KwData{Population $P_g$, Diversity threshold $\theta$, Diversity action (`mutate' or `replace')}
\KwResult{Diversified population}

Calculate diversity $D$ of $P_g$\;
\If{$D < \theta$}{
    \If{action == `mutate'}{
        Apply higher variance mutations to all individuals in $P_g$\;
    }
    \ElseIf{action == `replace'}{
        Replace a percentage of the population with new random individuals\;
    }
    Re-evaluate fitness for the affected individuals\;
}
\Return{$P_g$}\;
\end{algorithm}


\begin{algorithm}
\caption{Feedback Mechanism for Parameter Adjustment}
\label{alg:feedback_mechanism}
\KwData{Generation $g$, Average fitness $\overline{F}_g$, Previous average fitness $\overline{F}_{g-I}$, Threshold $\delta$, Increment $\Delta p$}
\KwResult{Adjusted mutation and crossover probabilities}

Calculate fitness improvement $\Delta \overline{F} = \overline{F}_g - \overline{F}_{g-I}$\;
\If{$\Delta \overline{F} < \delta$}{
    Increase $p_{\text{mut}}$ and $p_{\text{cross}}$ by $\Delta p$\;
}
\Else{
    Decrease $p_{\text{mut}}$ and $p_{\text{cross}}$ by $\Delta p$\;
}
\end{algorithm}


\begin{algorithm}
\caption{Early Stopping Criterion}
\label{alg:early_stopping}
\KwData{No improvement counter $no\_improvement$, Maximum allowed $R$ rounds}
\KwResult{Termination of the algorithm}

\If{No significant improvement in the last $R$ generations}{
    Increment $no\_improvement$\;
    \If{$no\_improvement \geq R$}{
        \textbf{Terminate the algorithm}\;
    }
}
\Else{
    Reset $no\_improvement$ to $0$\;
}
\end{algorithm}

\end{document}